\documentclass[letterpaper,twocolumn,preprintnumbers,amsmath,amssymb,amsfonts,floatfix,nofootinbib,noshowkeys]{revtex4}
\pdfoutput=1        
\usepackage{graphicx}
\usepackage{amssymb}
\usepackage{amsmath}
\usepackage{hypernat}
\usepackage{color}
\usepackage[utf8]{inputenc} 
\newif\ifcomment
\newif\ifprint
\newif\ifextra
\newif\ifcode
\codetrue
\graphicspath{{./pics/}}
\ifprint
 \codefalse
 \extrafalse
 \usepackage[linkbordercolor={0 0 .8},%
             citebordercolor={.8 0 0},%
             urlbordercolor={.8 0 .8},%
             raiselinks=true,%
             pdfborder={0 0 1 [3]}]{hyperref}
\else
 \usepackage[colorlinks,
             linkcolor=blue,citecolor=blue,anchorcolor=blue,urlcolor=red]{hyperref}
\fi
\ifpdf
\fi
\newcommand{\snn}         {\ensuremath{\sqrt{s_{\scriptscriptstyle{{\rm NN}}}}}}
\newcommand{\signn}       {\ensuremath{\sigma_{\scriptscriptstyle{{\rm NN}}}}}
\newcommand{\sigppb}      {\ensuremath{\sigma_{\scriptscriptstyle{{\rm pPb}}}}}
\newcommand{\sigpbpb}     {\ensuremath{\sigma_{\scriptscriptstyle{{\rm PbPb}}}}}
\newcommand{\sigauau}     {\ensuremath{\sigma_{\scriptscriptstyle{{\rm AuAu}}}}}

\newcommand{\Ncoll}       {\ensuremath{N_{\rm coll}}}
\newcommand{\Npart}       {\ensuremath{N_{\rm part}}}
\newcommand{\Nc}          {\ensuremath{N_{\rm c}}}
\newcommand{\Ncpart}      {\ensuremath{N_{\rm cpart}}}
\newcommand{\Nccoll}      {\ensuremath{N_{\rm ccoll}}}
\newcommand{\pT}          {\ensuremath{p_{\rm T}}}
\newcommand{\sigcc}       {\ensuremath{\sigma_{\rm cc}}}
\newcommand{\av}[1]       {\left<#1\right>}

\newcommand{\hrefurl}[1]  {\href{#1}{\url{#1}}}
\newcommand{\Eq}[1]       {Eq.~\ref{#1}}
\newcommand{\Ref}[1]      {Ref.~\cite{#1}}

\newcommand{\Fig}[1]      {Fig.~\ref{#1}}
\newcommand{\Figure}[1]   {Figure~\ref{#1}}
\newcommand{\Tab}[1]      {Tab.~\ref{#1}}
\newcommand{\Sec}[1]      {Sec.~\ref{#1}}
\newcommand{\Section}[1]  {Section.~\ref{#1}}
\newcommand{\App}[1]      {App.~\ref{#1}}

\newcommand{\gsim}        {\,{\buildrel > \over {_\sim}}\,}
\newcommand{\co}[1]       {}
\begin{document}
\title{Glauber modeling of high-energy nuclear collisions at sub-nucleon level}
\author{C.\ Loizides$^1$}
\affiliation{$^1$Lawrence Berkeley National Laboratory, Berkeley, California, 94720, USA}
\begin{abstract}\noindent
Glauber models based on nucleon--nucleon interactions are commonly used to characterize the initial state in high-energy nuclear collisions, and the dependence of its properties on impact parameter or number of participating nucleons.
In this paper, an extension to the Glauber model is presented, which accounts for an arbitrary number of effective sub-nucleon degrees of freedom, or active constituents, in the nucleons.
Properties of the initial state, such as the number of constituent participants and collisions, as well as eccentricity and triangularity, are calculated and systematically compared for different assumptions of how to distribute the sub-nuclear degrees of freedom and for various collision systems.
It is demonstrated that at high collision energy the number of produced particles scales with an average number of sub-nucleon degrees of freedom of between $3$ and $5$. 
The source codes for the constituent Monte Carlo Glauber extension, as well as for the calculation of the overlap area and participant density in a standard Glauber model, are made publicly available.
\end{abstract}
\maketitle
\section{Introduction}
\label{sec:intro}
Properties of the initial state in high-energy nuclear collisions are commonly calculated using a Glauber model~\cite{Miller:2007ri}.
In these calculations, nuclei are composed out of a set of nucleons, and the nuclear reaction is approximated by successive independent nucleon--nucleon~(NN) interactions assuming the nucleons travel in a straight line along the beam axis~(eikonal approximation). 
The so called ``optical'' Glauber calculations~\cite{Bialas:1976ed,Eskola:1988yh} assume a smooth matter density distribution for the makeup of the nuclei, while the Monte Carlo~(MC) based models~\cite{Alver:2008aq,Rybczynski:2013yba} distribute individual nucleons event-by-event, and collision properties are obtained by averaging over multiple events. 
In both cases, one usually uses a Fermi distribution for the radial direction and a uniform distribution for the solid angle.

These calculations can easily be extended to the sub-nucleon level by taking into account three valence quarks per nucleon in the collision process. 
It has recently been shown~\cite{Eremin:2003qn,Nouicer:2006pr,Adler:2013aqf,Adare:2015bua,Lacey:2016hqy,zheng} that particle production at mid-rapidity in high-energy nucleus--nucleus collisions scales almost linearly with the number of quark participants, without the need to introduce a contribution from a hard-scattering component scaling with the number of binary nucleon--nucleon collisions.
Further interest in such calculations arises since understanding the observed azimuthal momentum anisotropy as a result of anisotropic pressure gradients formed early-on due to the spatial anisotropy of the initial state in pA and even pp collisions~(see \Ref{Loizides:2016tew} for a recent summary) needs calculations of the initial state in small systems at the sub-nucleon level~\cite{Bozek:2016kpf}.

In this paper, an extension of the MC Glauber model is presented, which generalizes the collision process by accounting for an arbitrary, but fixed, number of effective sub-nucleon degrees of freedom, or active constituents, in the nucleons.
This description can obviously not account for the partonic structure of a nucleon, which depends on the momentum transfer~($Q^2$) and fraction of nucleon momentum~(Bjorken-$x$)\co{ carried by the partons}.
However, \co{assuming soft particle production is dominated by a generic additive source,} the constituent MC Glauber calculation can be used to effectively model the average number of active degrees of freedom, which contribute to soft particle production, and to study the dependence on collision energy and species.
In \Sec{sec:calculation} the standard MC Glauber model is briefly recalled, while in \Sec{sec:extension} its extension to the sub-nucleon level is discussed.
\Section{sec:results} discusses properties of the initial state, such as the number of constituent participants and collisions, as well as eccentricity and triangularity, calculated for a variety of different assumptions to distribute the sub-nuclear degrees of freedom and for various collision systems.
\Section{sec:sum} provides a short summary.
The code for the constituent MC Glauber program is described in \App{sec:code}.
Additional calculations of the overlap area and participant density are discussed in \App{sec:area}.

\section{MC Glauber calculation}
\label{sec:calculation}
The Glauber calculation of a nucleus--nucleus collision is done as described in \Ref{Alver:2008zza}. 
First, the positions of each of the $A$ nucleons in a nucleus are determined according to the measured charge density distribution of the nucleus extracted from low-energy electron scattering experiments~\cite{DeJager:1987qc}. 
For spherical nuclei, such as Pb, the distribution is taken to be uniform in azimuthal and polar angles, and a two-parameter Fermi function
\begin{equation}\label{eq:1}
  \rho(r)=\rho_0 \left(1+\exp\left(\frac{r-R}{a}\right)\right)^{-1}
\end{equation} 
in the radial direction.
In \Eq{eq:1}, $R$ is the nuclear radius, and $a$ is the skin depth, and the overall normalization $\rho_0$ is not relevant for the calculation. 
To mimic a hard-core repulsion potential in the context of the MC Glauber model, one usually requires a minimum inter-nucleon separation~($d_{\rm min}$) of $0.4$~fm between the centers of the nucleons.
These excluded-volume effects of the nucleons distort the resulting nuclear density and can be absorbed by rescaling the charge-density parameters\co{ to $R=6.65$ and $a=0.46$~fm}~\cite{Shou:2014eya}.
The standard and rescaled values for Au and Pb nuclei are given in \Tab{tab:1}; for other nuclei see \Ref{Loizides:2014vua}.

\begin{table}[t]
\begin{center}
  \begin{tabular}{l|c|c|c|c}
    Nucleus       & $R$~(fm)      & $R_{\rm s}$ (fm) &  $a$~(fm) & $a_{\rm s}$~(fm) \\
    \hline
    \hline
    ${}^{197}$Au   & $6.38\pm0.13$ & 6.42 & $0.535\pm0.053$ & 0.44\\
    ${}^{208}$Pb   & $6.62\pm0.06$ & 6.65 & $0.546\pm0.010$ & 0.46\\
  \end{tabular}
  \caption{\label{tab:1}Standard and rescaled charge-density parameters\co{ for Au and Pb nuclei}.}
\end{center}  
\end{table}
\begin{table}[t!]
\begin{center}
  \begin{tabular}{l||c|c|c|c|c|c}
    $\sqrt{s}$ (TeV) & 0.019 & 0.2  & 2.76 & 5.02  & 7    & 13 \\
\hline
    $\signn$ (mb)    & 33    & 42  & 64    & 70   & 74    & 78   \\
\hline
\hline
           $\Nc$     &\multicolumn{6}{c}{$\sigcc$ (mb)} \\     
\hline
               3      & 6.3   & 9.2 & 18.3  & 21.1 & 23.0  & 25.2 \\
               3$^{*}$& 5.8    & 8.1 & 15.5  & 17.9 & 19.7  & 21.6  \\
               5      & 2.4   & 3.6 & 8.4   & 10.3 & 11.4  & 12.7 \\
               7      & 1.2   & 1.9 & 4.6   & 5.7  & 6.5   & 7.4  \\
              10      & 0.6   & 0.9 & 2.2   & 2.8  & 3.3   & 3.8  \\
              20      & 0.1   & 0.2 & 0.5   & 0.6  & 0.7   & 0.8  \\
  \end{tabular}
  \caption{\label{tab:2}Values used for $\signn$ at various $\snn$ at nucleon level, as well as corresponding $\Nc$ and $\sigcc$ parameters at sub-nucleon level. The modified case is indicated with $^{*}$~(see text).}
\end{center}  
\end{table}

Second, the collision impact parameter~($b$) is determined from ${\rm d}N/{\rm d}b \propto b$, and the centers of the nuclei are shifted to $(-b/2,0,0)$  and $(b/2,0,0)$~\footnote{The reaction plane, i.e.\ the plane defined by the impact parameter and the beam direction, is given by the $x$- and $z$-axes, while the transverse plane is given by the $x$- and $y$-axes.}.
Following the eikonal ansatz, the nucleons are assumed to move along a straight trajectory along the beam axis.
Their transverse positions are held constant during the short passage time of the two high-energy nuclei, while their longitudinal coordinate does not play a role in the calculation.
The nuclear reaction is modeled by successive independent interactions between two nucleons from different nuclei.
The interaction strength between two nucleons is parameterized by the nucleon--nucleon inelastic cross section~($\signn$).
Two nucleons from different nuclei are supposed to collide if their relative transverse distance is less than 
\begin{equation}
  \label{eq:2}
  D = \sqrt{\signn/\pi}\,.
\end{equation}
A nucleus--nucleus collision is accepted if at least one such nucleon--nucleon collision was obtained.

\begin{figure}[t]
\begin{center}
   \includegraphics[width=0.48\textwidth]{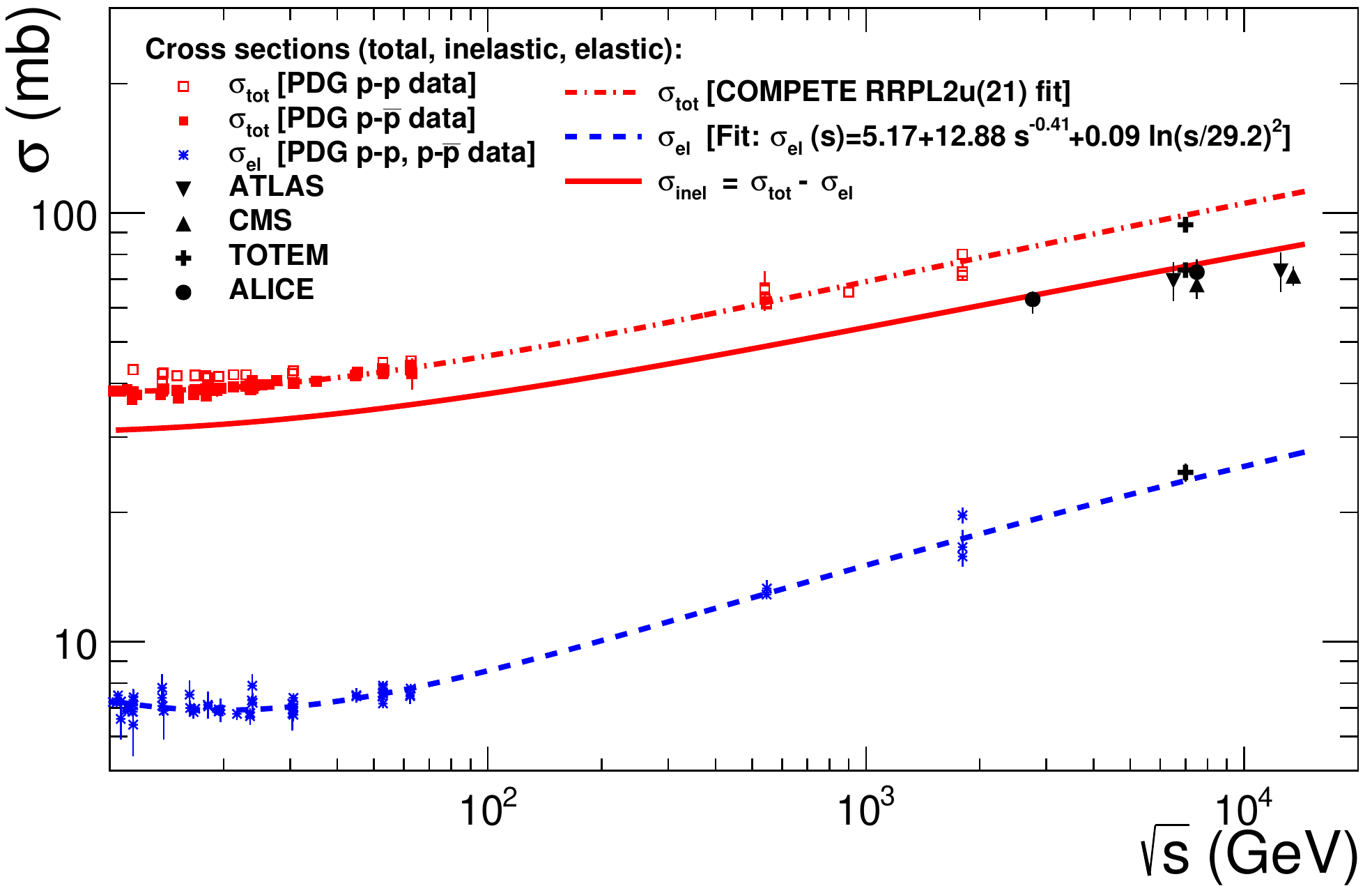}
   \caption{\label{fig:compfit} Available data of total, elastic and inelastic cross sections measured in pp and p$\bar{\mathrm {p}}$ collisions~\cite{Agashe:2014kda,Aad:2011eu,Antchev:2011vs,Chatrchyan:2012nj}. The data~\cite{ATLAS-CONF-2015-038,CMS:2016ael} at 13 TeV are preliminary. The curves are fits performed by the COMPETE Collaboration~\cite{Cudell:2002xe}. The figure, originally from \cite{dde}, was adapted from \Ref{Abelev:2013qoq}.}
\end{center}
\end{figure}

The values used for $\signn$ are usually obtained from the difference of total and elastic pp cross section measurements~\cite{Agashe:2014kda,Aad:2011eu,Antchev:2011vs,Chatrchyan:2012nj},
or interpolated using fits performed by the COMPETE Collaboration~\cite{Cudell:2002xe} as shown in \Fig{fig:compfit}.
Common values of $\snn$ are summarized in \Tab{tab:2} for a number of collision energies, and in good agreement with the COMPETE fits.
At 13 TeV, however, the preliminary data~\cite{ATLAS-CONF-2015-038,CMS:2016ael} indicate that the fit overpredicts the cross section by about $15$\%.
As a compromise, $78$~mb, which is between the central value of the data and the fit, and roughly within $1\sigma$ of the experimental uncertainty, is given in \Tab{tab:2}, and used in the following.

To estimate systematic uncertainties for calculated quantities\co{~(like $\Npart$)} it is suggested to systematically modify the parameters of the calculation~\cite{Alver:2008zza}.
One typically varies the parameters of the nuclear density profile within the measured $1\sigma$ uncertainties, the minimum inter-nucleon separation distance by 100\%, and the $\signn$ by about $\pm$3~mb and $\pm$5~mb at the Relativistic Heavy Ion Collider~(RHIC) and the Large Hadron Collider~(LHC), respectively.

\begin{figure}[t]
\begin{center}
   \includegraphics[width=0.48\textwidth]{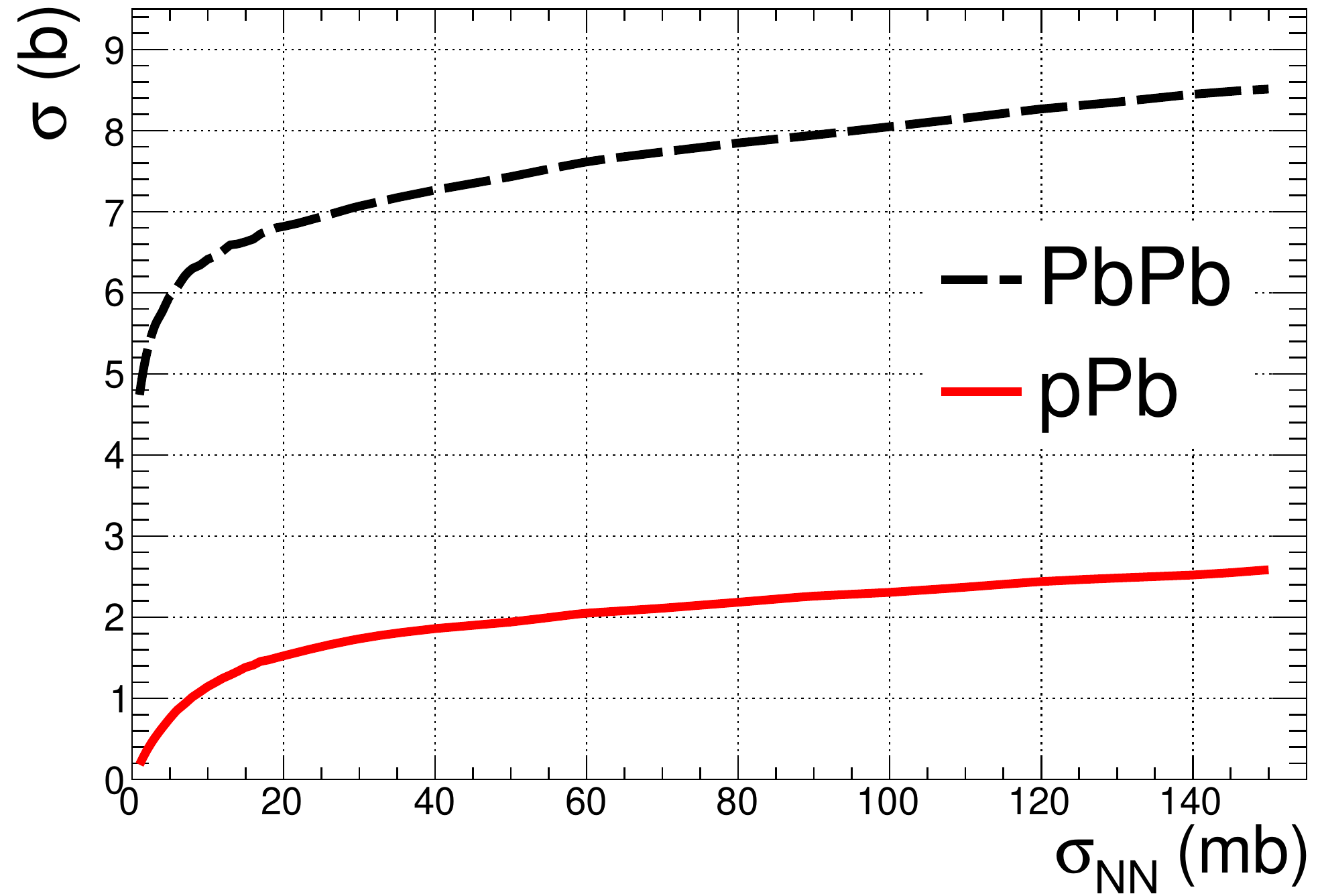}
   \caption{Calculated total cross sections for PbPb and pPb collisions as a function of $\signn$.
     \label{fig:xsec}}
\end{center}
\end{figure}

The Glauber calculation gives $\sigpbpb^{\rm MC}=7.6\pm0.2$~b and $\sigppb^{\rm MC}=2.1\pm0.1$~b for the total PbPb and pPb cross sections, in good agreement with the measured values of $\sigpbpb=7.7\pm0.6$~b at $\snn=2.76$~TeV~\cite{ALICE:2012aa} and $\sigppb=2.06\pm0.08$~b at $\snn=5.02$~TeV~\cite{Khachatryan:2015zaa}, respectively.
For PbPb at $\snn=5.02$ a total cross section of $\sigpbpb^{\rm MC}=7.7\pm0.2$~b is predicted.
The total cross sections of PbPb and pPb as a function of $\signn$ are shown in \Fig{fig:xsec} calculated using the central values of the parameters (i.e.\ without systematic uncertainties, which would be about 3 and 8\%, respectively).

\begin{figure}[t]
\begin{center}
   \includegraphics[width=0.235\textwidth]{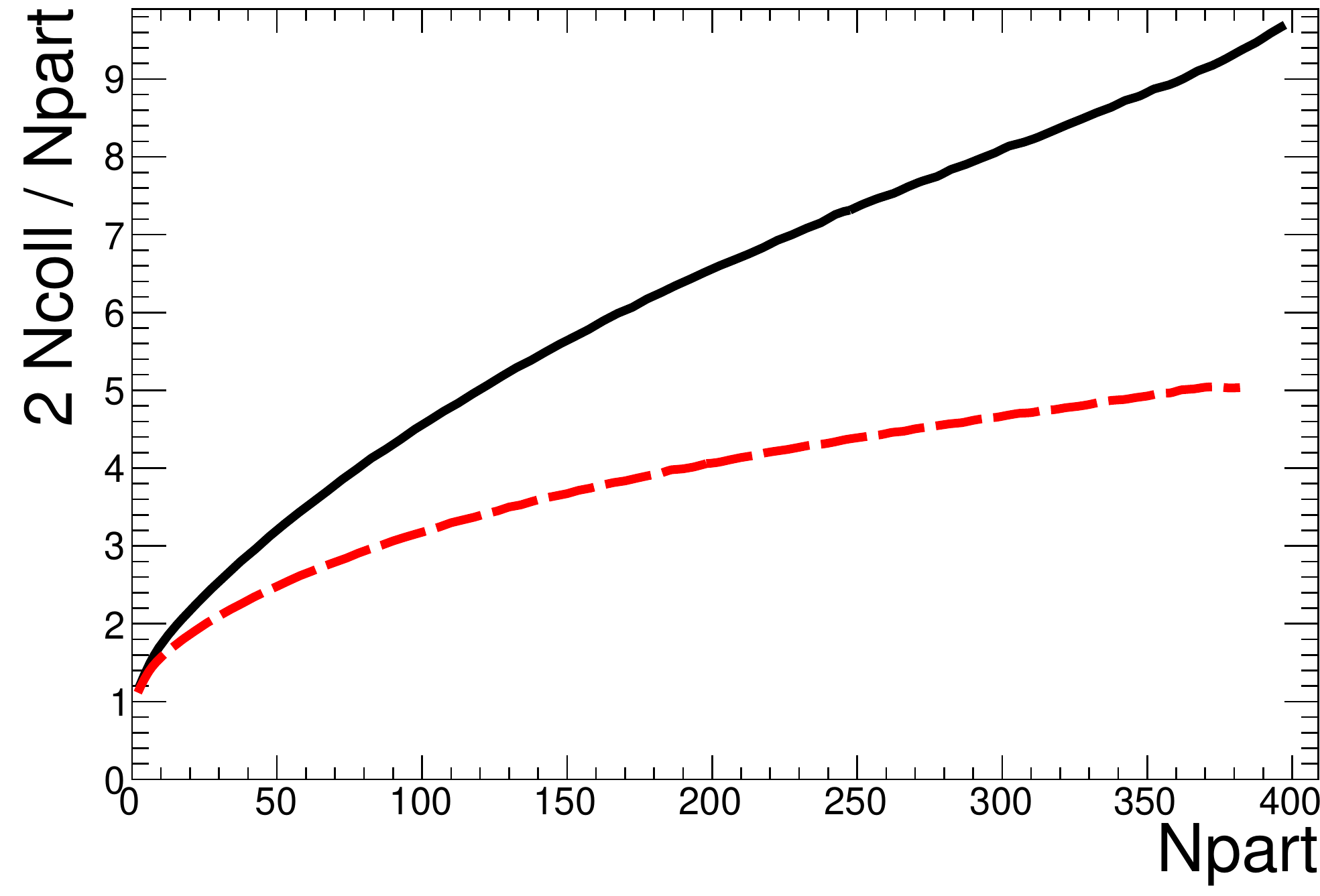}
   \includegraphics[width=0.235\textwidth]{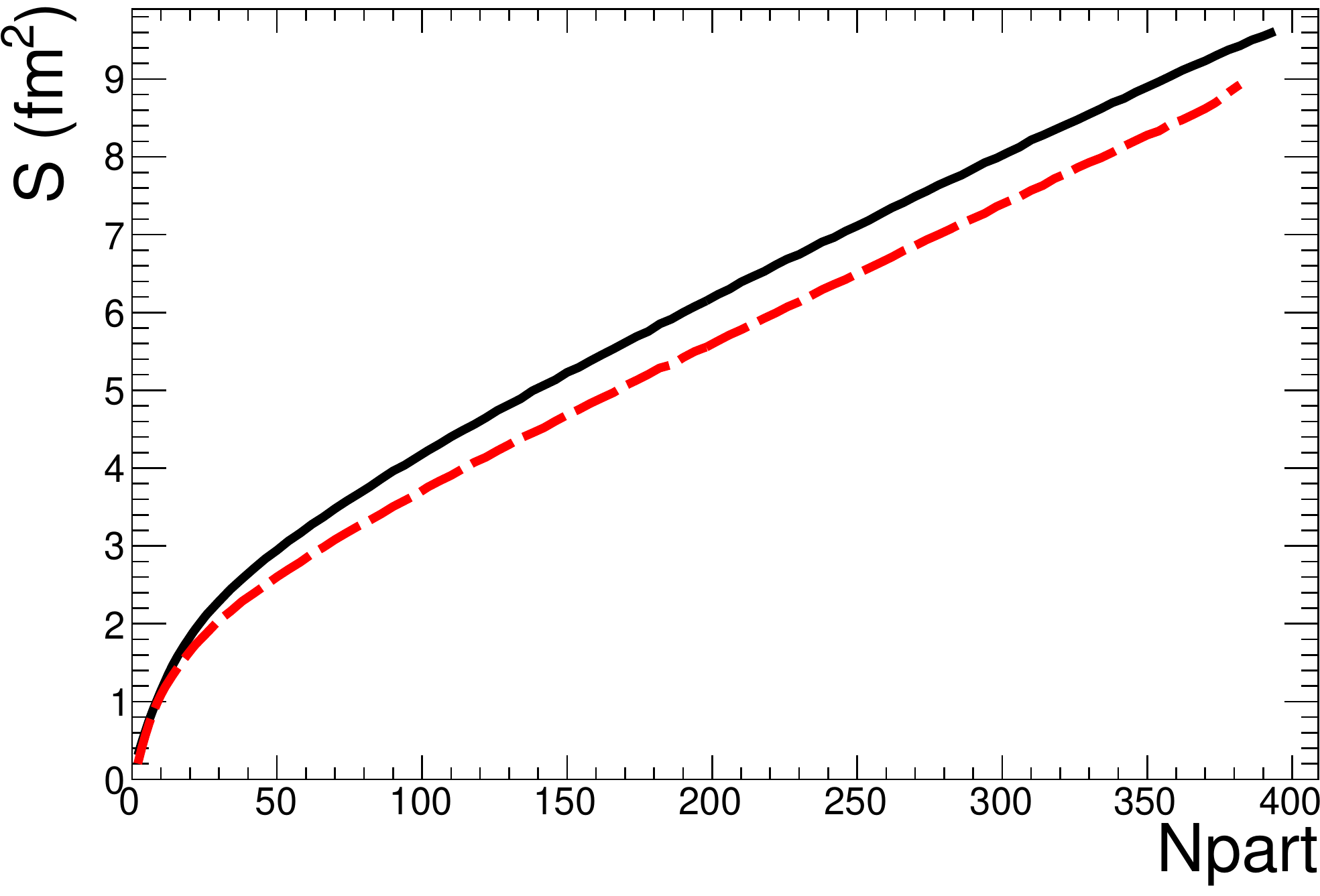}
   \includegraphics[width=0.235\textwidth]{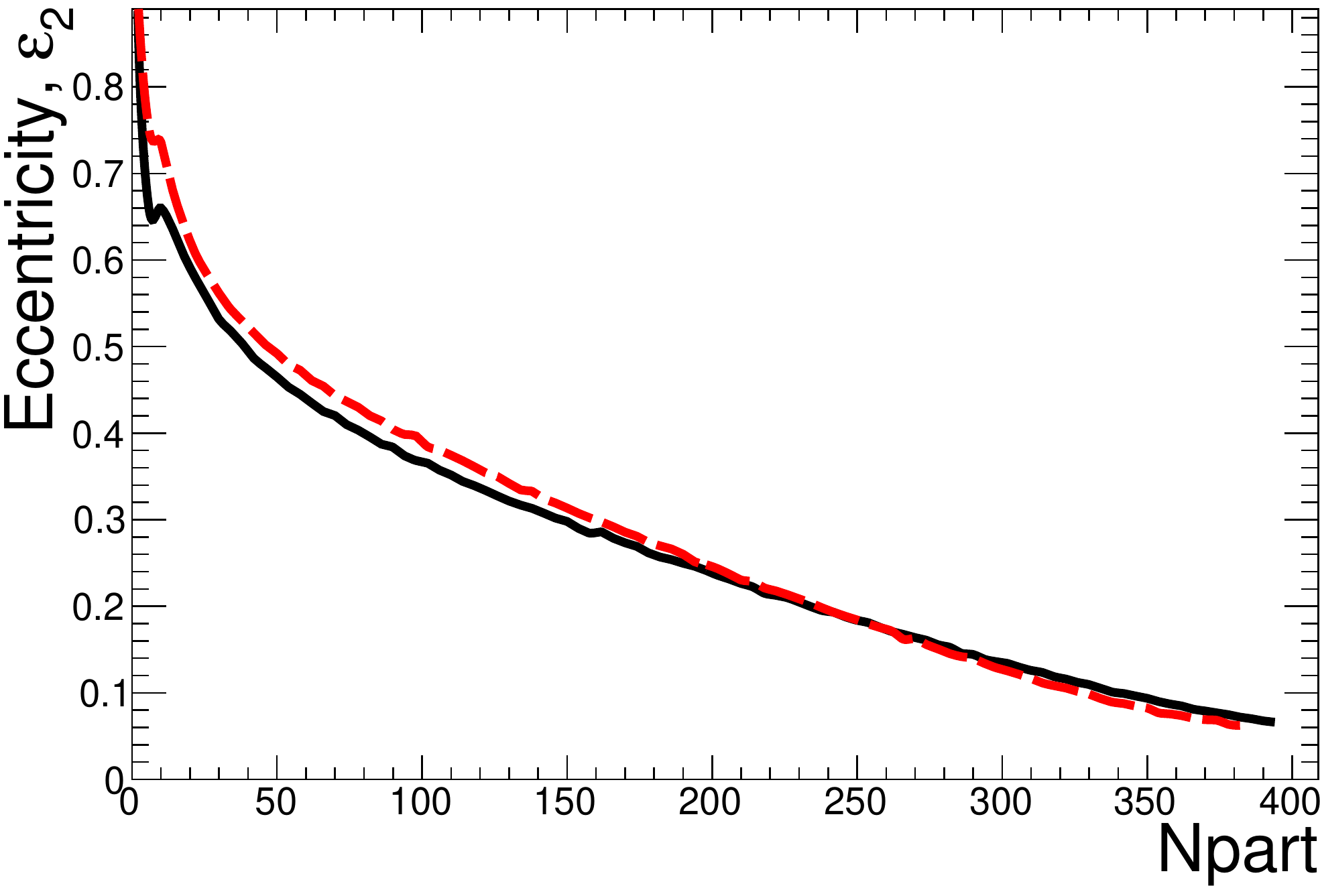}
   \includegraphics[width=0.235\textwidth]{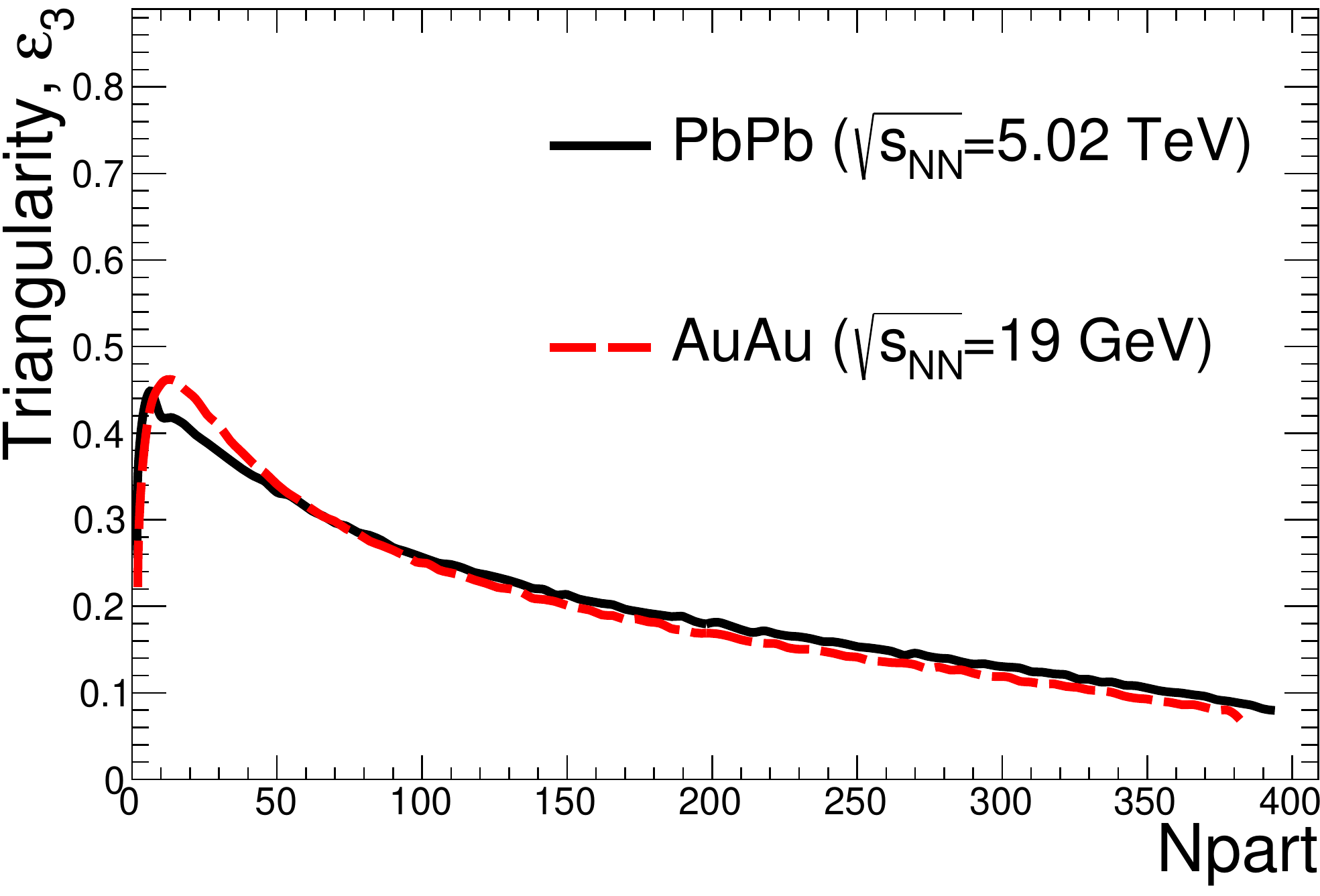}
   \caption{\label{fig:propg}Geometric properties ($2\Ncoll\Npart$, $S$, $\varepsilon_2$ and $\varepsilon_3$ from top left to bottom right panels) computed with Glauber MC for AuAu collisions at $\snn=19$ GeV and PbPb collisions at $\snn=5.02$ TeV.}
\end{center}
\end{figure}

MC Glauber calculations are typically used to compute geometrical properties of the collision, such as the number of participating nucleons in the collision, $\Npart$, i.e.\ the number of nucleons that are hit at least once, or the number of independent nucleon--nucleon collisions, $\Ncoll$, i.e.\ the total number of collisions between nucleons.
Particle production at low $\pT$ roughly scales with $\Npart$~\cite{Alver:2010ck}, while hard processes in the absence of strong final state modification scale with $\Ncoll$~\cite{Aggarwal:2000th,Adler:2005ig,Chatrchyan:2012vq}.

Examples of geometrical properties are shown in \Fig{fig:propg}, and have been discussed extensively in the literature~(e.g.\ see \Ref{Alver:2008zza}).
The ratio between $\Ncoll/\Npart$ normalized to that of pp (i.e.\ $1/2$), which has been argued to be a measure for the relative importance of hard versus soft processes, rises with centrality
and in particular with collision energy.
The overlap area of the two colliding nuclei is proportional to $S=\sqrt{\sigma^2_{x}\sigma^2_{y}-\sigma^2_{xy}}$, given by the \mbox{(co-)}variances of the participant distributions in the transverse plane~\cite{Alver:2008zza}.
The area can also be directly computed from the MC as explained in \App{sec:area}, leading to a slightly different shape for peripheral collisions.
The eccentricity~\cite{Alver:2006wh} and triangularity~\cite{Alver:2010gr} of the collision region, given by $\varepsilon_{i}=\left<r^i\cos(i\phi-i\psi_i)\right>/\left<r^i\right>$~(for $i=2$ and $3$, respectively)~\cite{Teaney:2010vd}, 
are used to characterize the initial geometrical shape. 
They are similar between AuAu and PbPb collisions, and at different collision energies.

\section{Extension to sub-nucleon level}
\label{sec:extension}
The calculation can be readily extended to the sub-nucleon level by assuming that a nucleon carries $\Nc$ degrees of freedom.
Often $\Nc=3$ for three constituent quarks~\cite{Eremin:2003qn,Adler:2013aqf,Adare:2015bua}, but larger numbers (up to $\Nc=17$) 
have previously~\cite{d'Enterria:2010hd} been used to account for the effective number of partonic degrees of freedom.
Generalizing \Eq{eq:2}, the interaction between two constituents can be modeled by an effective 
parton--parton cross section~($\sigcc$) in the same way as before, 
i.e.\ two constituents from different nuclei collide if their relative transverse distance is less than 
\begin{equation}
  \label{eq:3}
  D = \sqrt{\sigcc/\pi}\,.
\end{equation}
The hard-sphere approximation differs from the approach e.g.\ implemented in \Ref{Bozek:2016kpf} where a Gaussian shape is assumed for the partonic inelasticity profile.

\begin{figure}[t]
\begin{center}
   \includegraphics[width=0.45\textwidth]{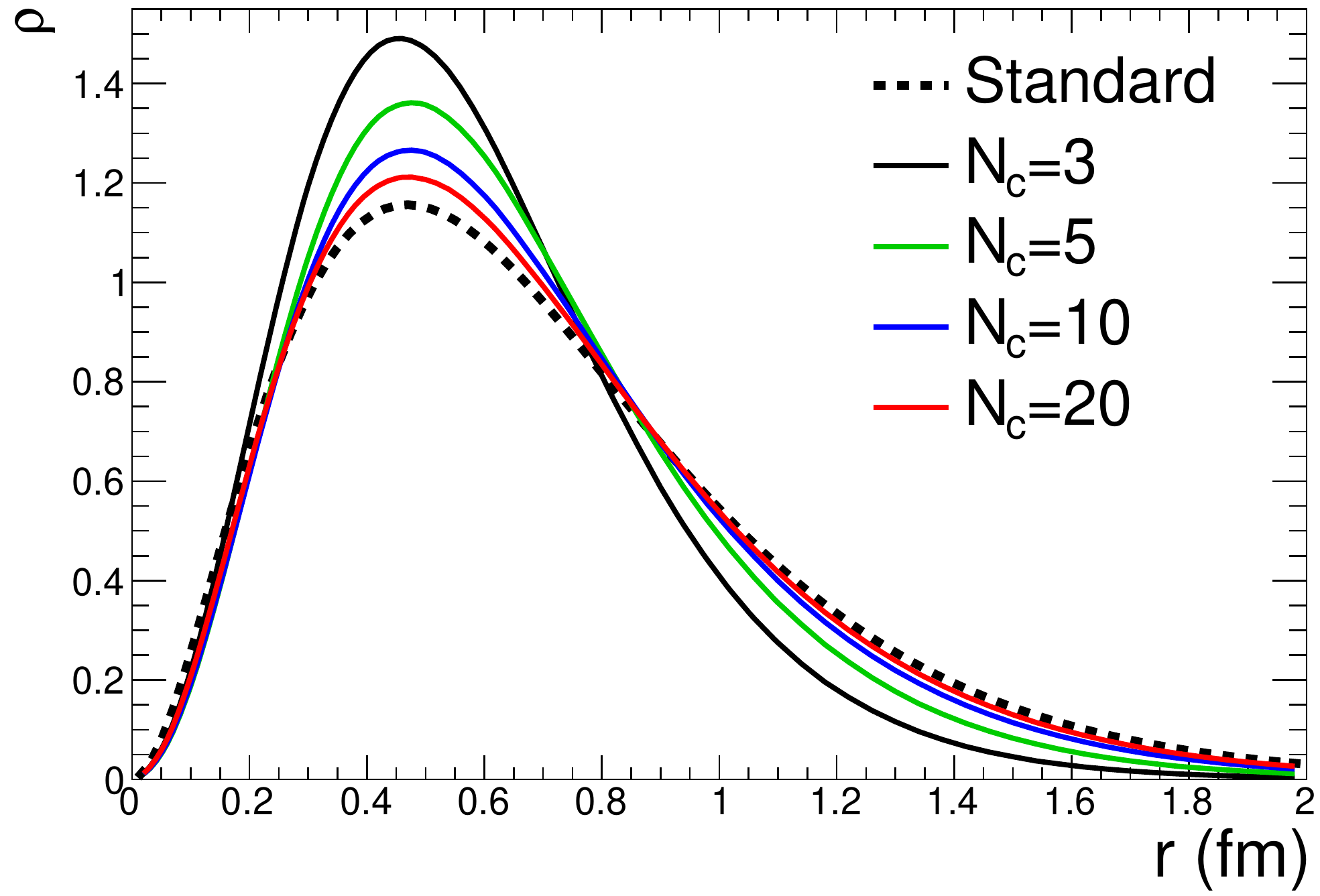}
   \caption{\label{fig:pp1}Radial distribution of constituents after recentering when constructed from the standard parametrization (\Eq{eq:4}) 
     for different $\Nc$.}
\end{center}
\end{figure}
\begin{figure}[t]
\begin{center}
   \includegraphics[width=0.45\textwidth]{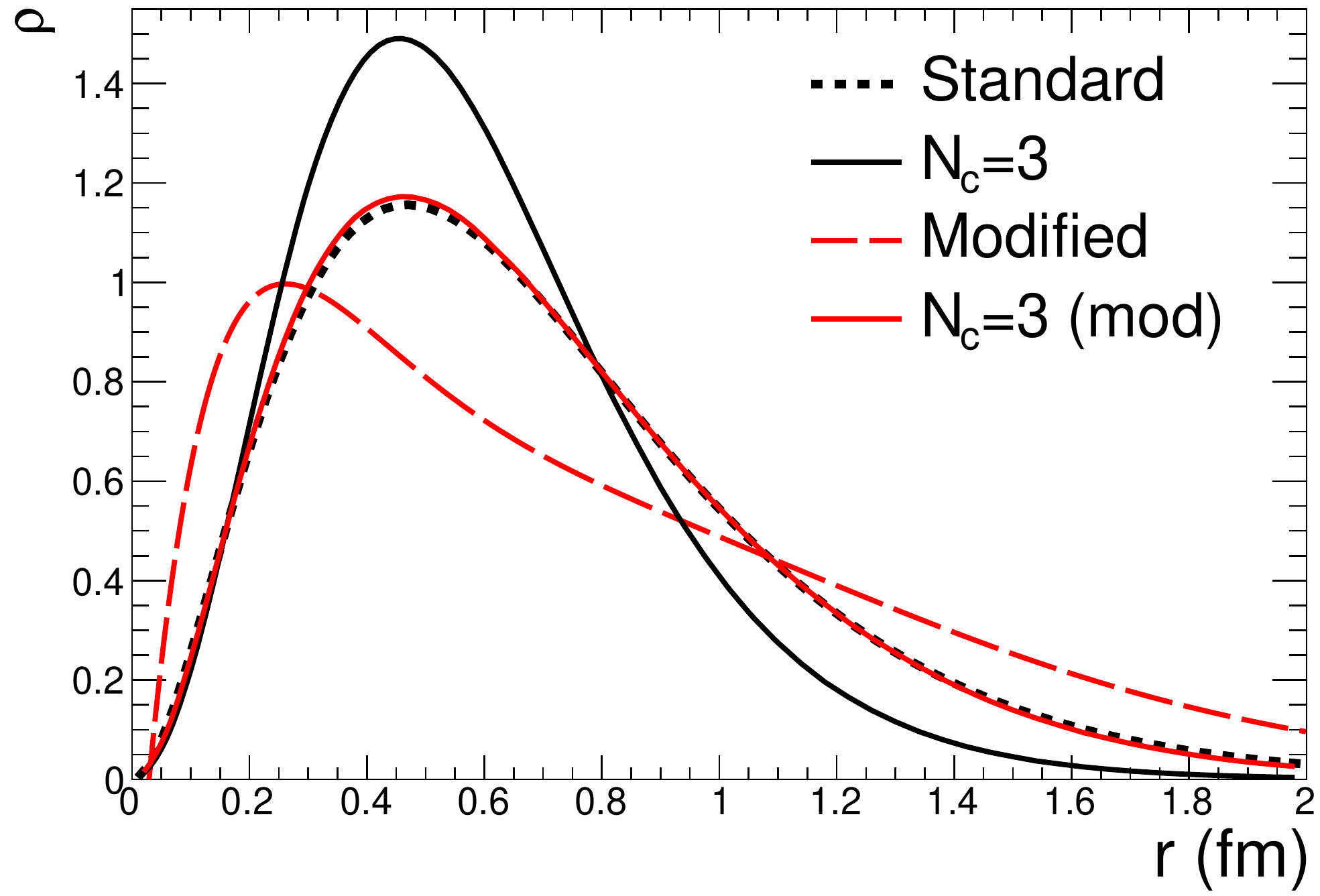}
   \caption{\label{fig:pp2}Radial distribution of constituents after recentering when constructed from the standard (\Eq{eq:4})
      or modified (\Eq{eq:5}) parametrizations for $\Nc=3$.}
\end{center}
\end{figure}

There are two, somewhat limiting, cases to distribute sub-nucleon degrees of freedom.
The first is to bind constituents to nucleons making up the nucleus~(labeled as ``bound'' in figures). 
In this case, $\Nc$ constituents are radially distributed centered around each nucleon according to 
\begin{equation}
  \label{eq:4}
  \rho(r) = \exp\left(-r/R\right)
\end{equation}
with $R=0.234$~fm based on the measured form factor of the proton~\cite{Hofstadter:1956qs}. 
The second is to freely distribute constituents over the whole nucleus~(labeled as ``free'' in figures). 
In this case $A\times\Nc$ constituents are distributed according to \Eq{eq:1}.~\footnote{This is conceptually similar to the optical approach used in~\cite{Eremin:2003qn}.}
In both cases, a hard core repulsion potential is not considered.

When the constituents are bound to nucleons, 
recentering of the constituents to align with the centers of their respective nucleons,
introduces a distortion of the resulting radial constituent distribution.
The effect is most dramatic for $\Nc=3$, and reduces quickly with increasing number of constituents as shown in \Fig{fig:pp1}.
For $\Nc=3$, the distortion can be avoided by distributing the constituents according to an empirically determined function~\cite{Mitchell:2016jio}
\begin{equation}
  \label{eq:5}
  \begin{aligned}
    \rho(r) = \,\, &  r^2\,\exp\left(-r/R\right) \times \left[(1.22-1.89r+2.03r^2)\right. \\ 
    & \left.(1+1/r-0.03/r^2)(1+0.15r)\right]
  \end{aligned}
\end{equation}
as shown in \Fig{fig:pp2}. Equation~\ref{eq:5} holds for $\Nc=3$. 
Hence, it is used as an alternative to \Eq{eq:4} only in the case of constraining 3 constituents to nucleons.
This case is labeled as ``mod'' when displayed in figures.  
\co{In general, and i}If not otherwise specified in the following, the constituents are not recentered.
In any case, not recentering has a negligible effect on the center of a nucleus since the deviations from the center-of-mass average out over $A\Nc$ degrees of freedom.

\begin{figure}[t]
\begin{center}
   \includegraphics[width=0.235\textwidth]{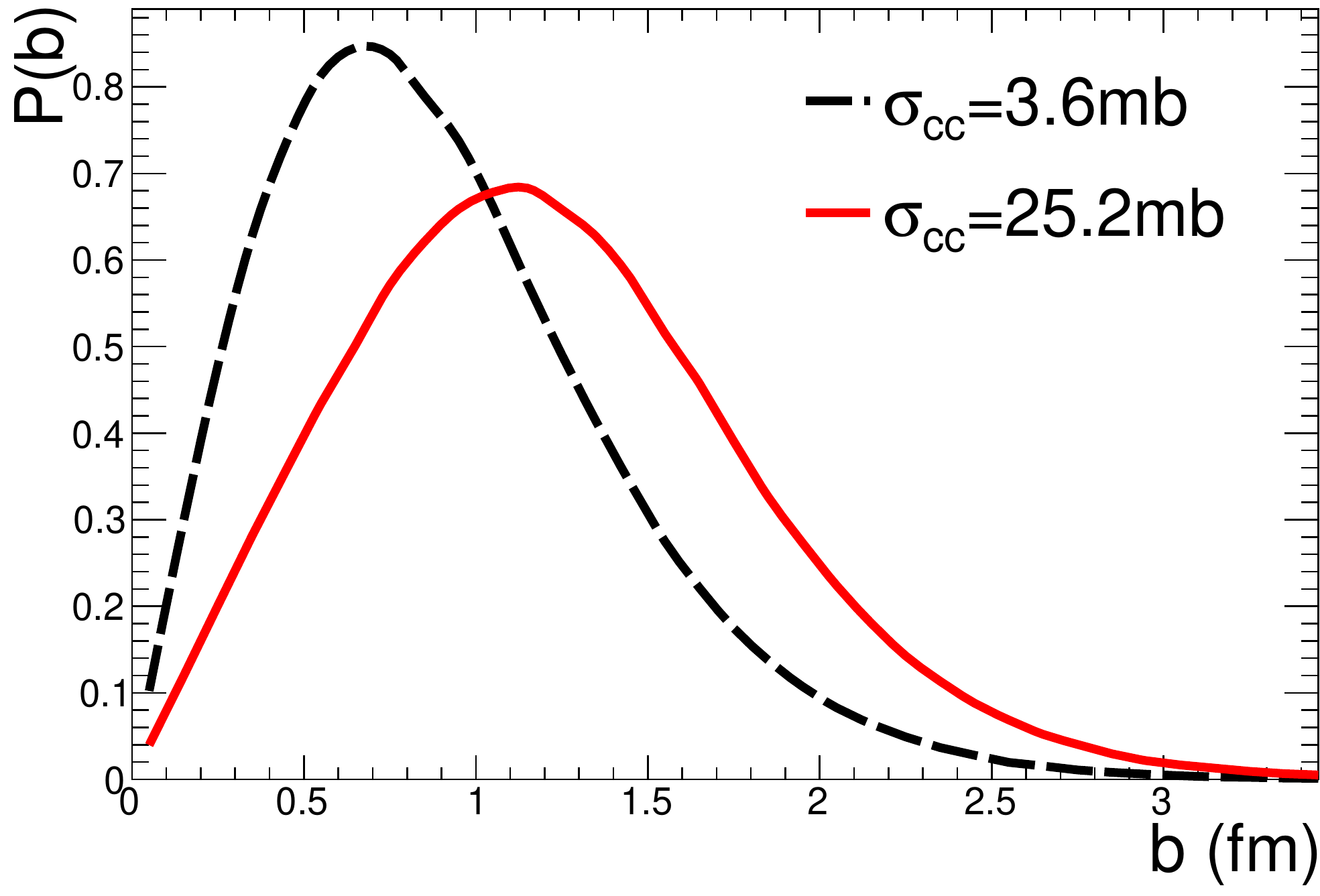}
   \includegraphics[width=0.235\textwidth]{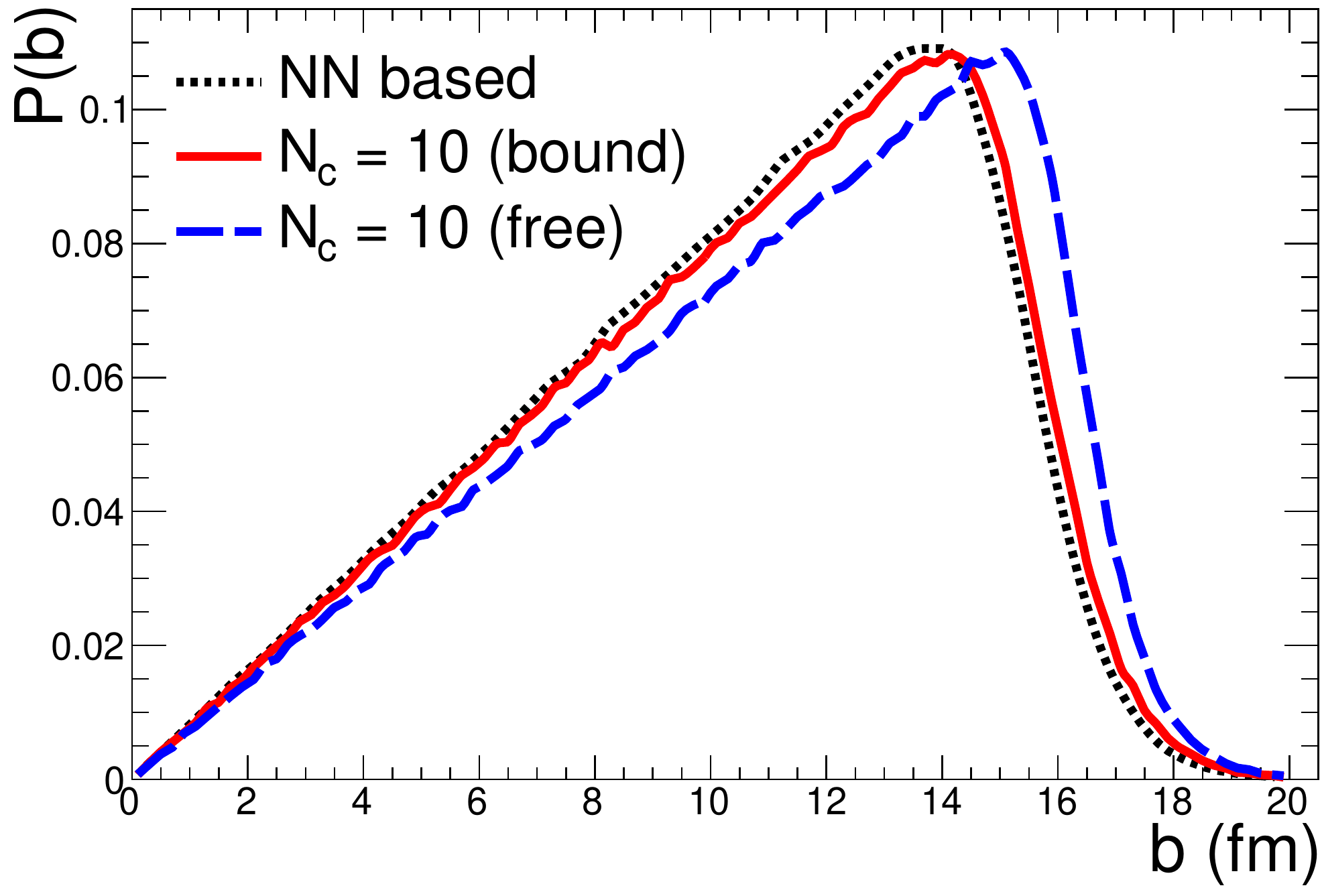}
   \caption{\label{fig:impb}Impact parameter distribution for pp (left) and PbPb (right) collisions. In case of pp, $\sigcc=3.6$ and $25.2$ mb with $\Nc=3$ are used for pp collisions at $\sqrt{s}=0.019$ and $13$ TeV, respectively. For PbPb collisions at $\snn=5.02$ TeV, the standard nucleon-based approach is compared to bound and freely-distributed cases using $\sigcc=3$ mb and $\Nc=10$.}
\end{center}
\end{figure}

The resulting impact parameter distributions differ from a straight line ($\propto b$) which holds in the case of a hard-sphere profile.
Examples are shown in \Fig{fig:impb} for pp collisions at $\sqrt{s}=0.019$ and $13$ TeV in the left, and for PbPb collisions at $\snn=5.02$ TeV in the right panel. 
In the case of pp, the distributions are obtained for $\Nc=3$ with $\sigcc=3.6$ and $25.2$ mb, and clearly extend beyond the hard-sphere limit of about $1.0$ and $1.6$~fm, respectively. 
In the case of PbPb, the distribution obtained for the standard NN based approach is compared to the bound and freely-distributed cases for $\sigcc=3$ mb and $\Nc=10$.
Freely-distributing constituents instead of binding them into nucleons generally leads to a wider impact parameter distribution.

\begin{figure}[t]
\begin{center}
   \includegraphics[width=0.45\textwidth]{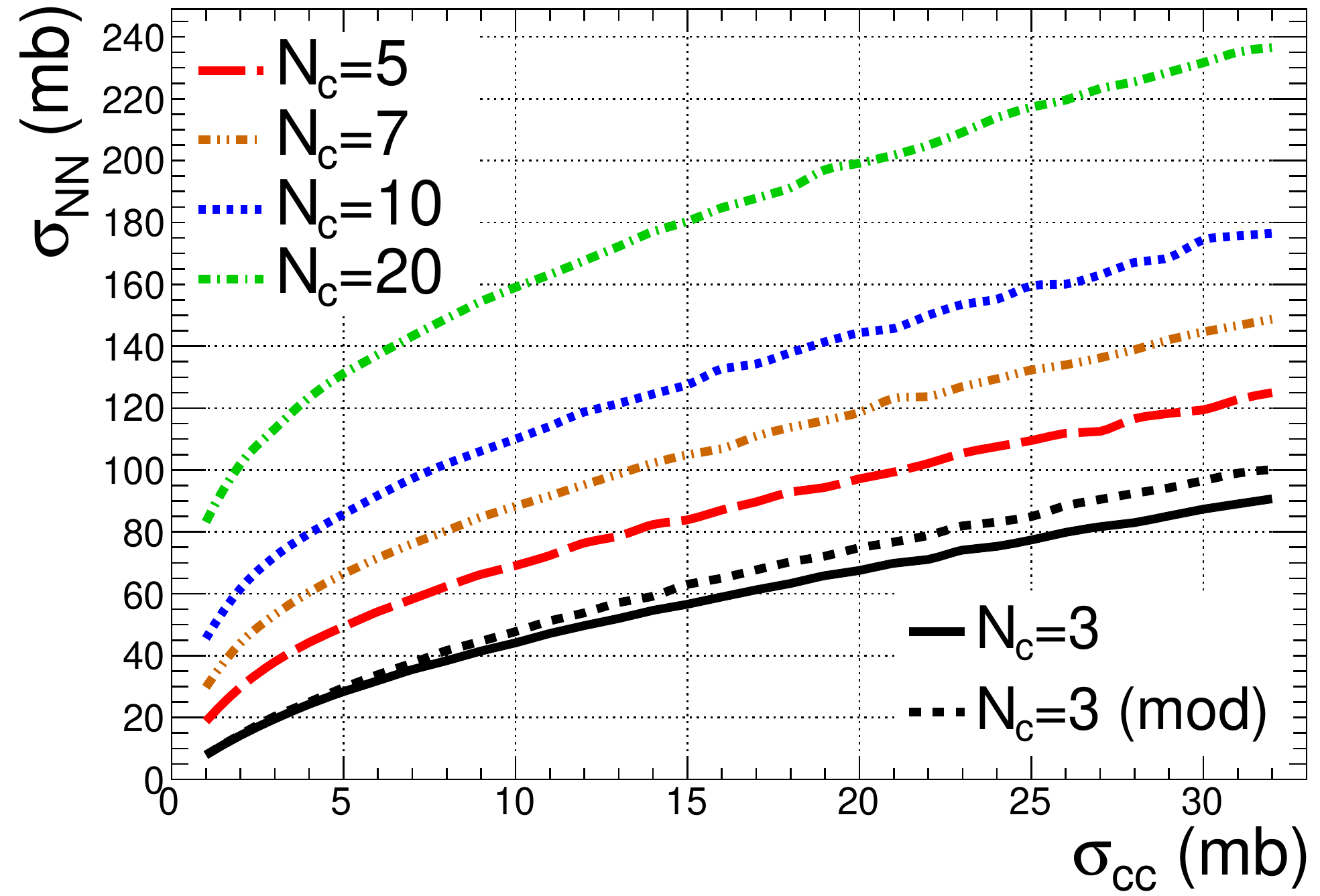}
   \caption{\label{fig:xsecpp}Calculated $\signn$ for various choices of $\Nc$ versus $\sigcc$. Parameters for commonly used $\signn$ are listed in \Tab{tab:2}.}
\end{center}
\end{figure}
\begin{figure}[t]
\begin{center}
   \includegraphics[width=0.45\textwidth]{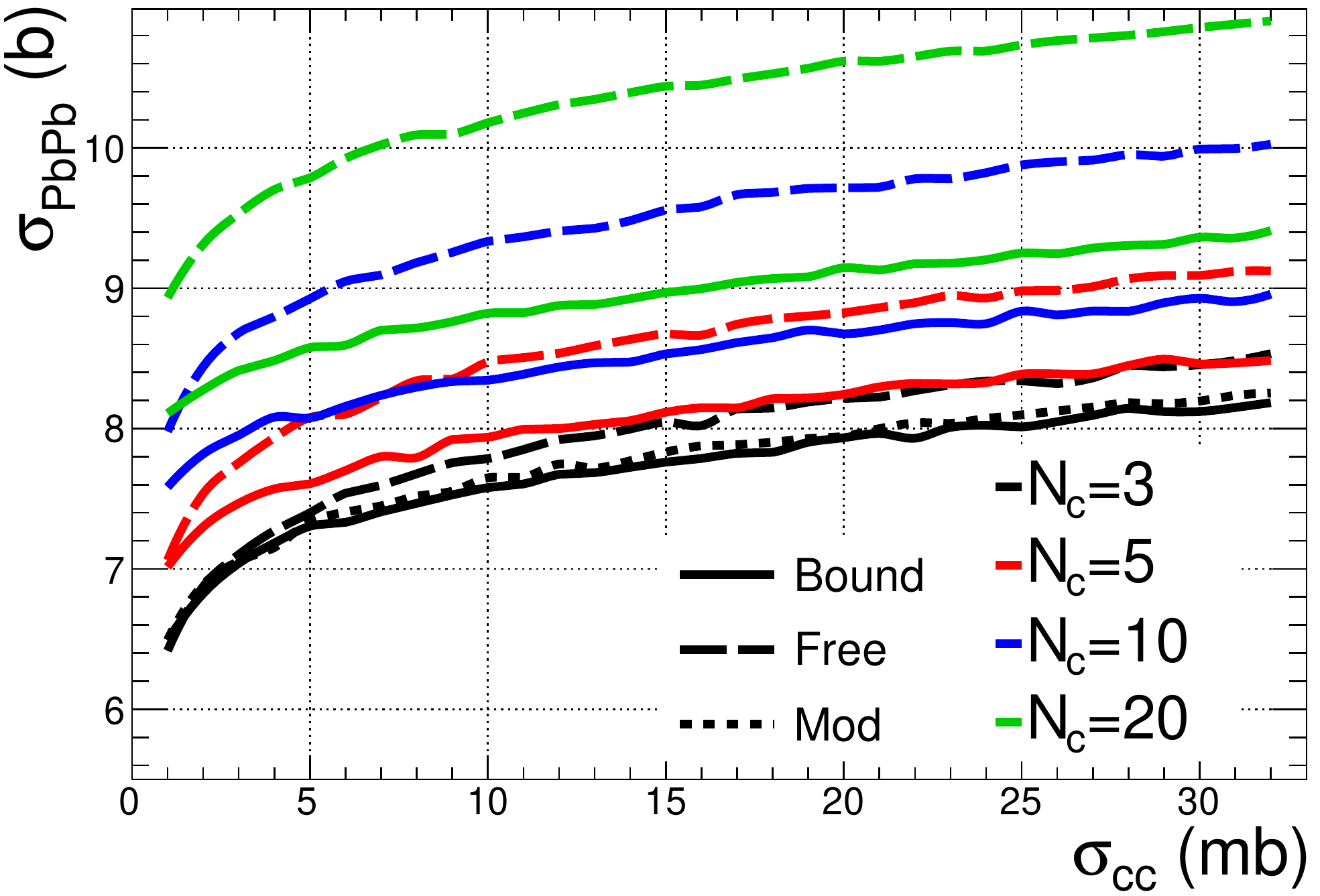}
   \includegraphics[width=0.45\textwidth]{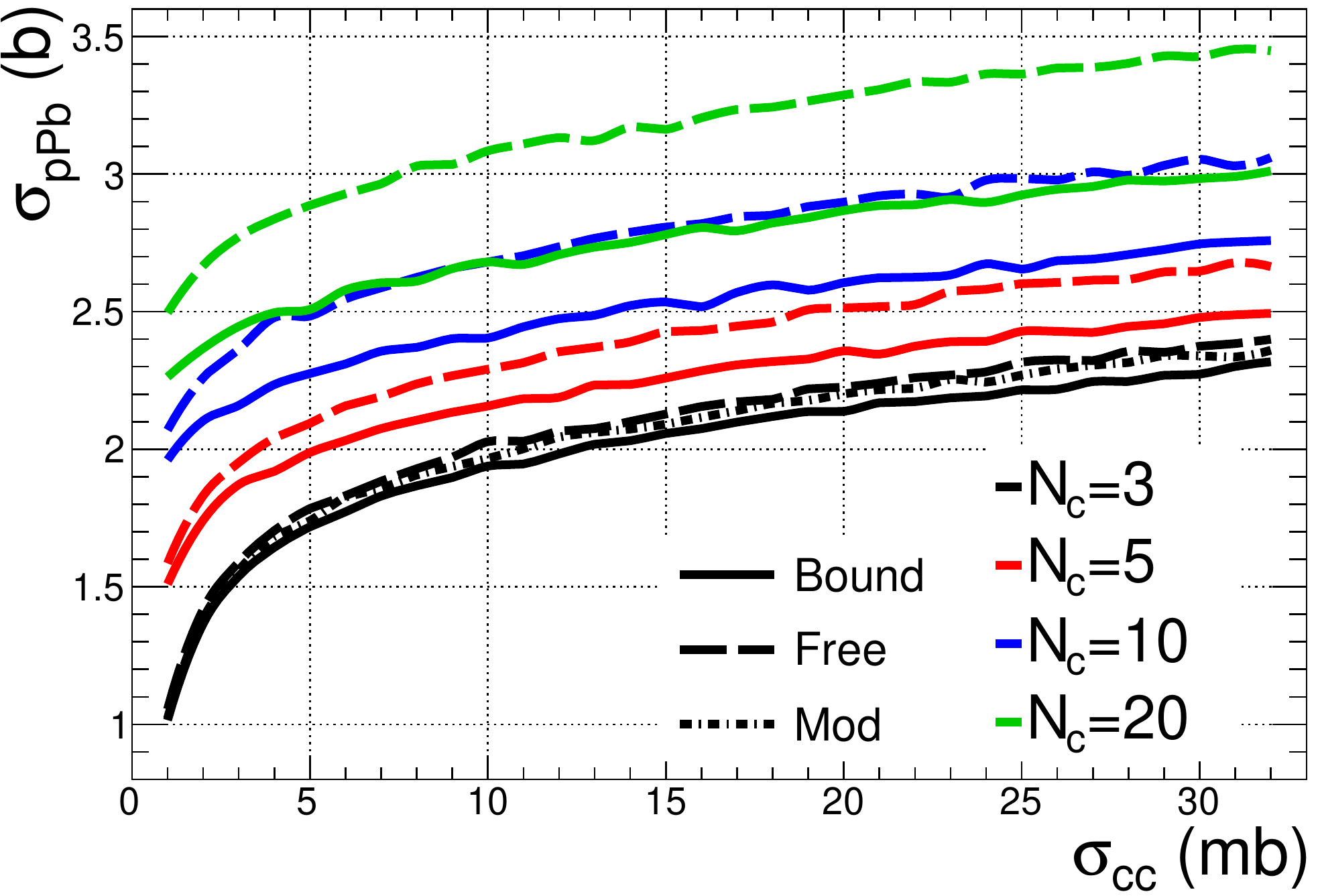}
   \caption{\label{fig:xsecpbpb}Calculated total cross section for PbPb (top) and pPb (bottom) collisions for various choices of $\Nc$ versus $\sigcc$ 
     for the bound and free cases.}
\end{center}
\end{figure}

One way to constrain the parameters of the calculation is to compare with nuclear reaction cross sections.
Nuclear reaction cross sections can be computed by counting if there was at least one collision among two constituents.
\Figure{fig:xsecpp} shows the dependence of $\signn$ on $\sigcc$ for various choices of $\Nc$.
As expected, $\signn$ strongly increases with increasing $\sigcc$ and $\Nc$.
For $\Nc=3$, the two parametrizations lead to a small but noticeable difference on $\signn$ for $\sigcc\gsim10$~mb.
Values for $\Nc$ and $\sigcc$ that correspond to commonly used $\snn$ are summarized in \Tab{tab:2}. 

\begin{figure}[t]
\begin{center}
   \includegraphics[width=0.42\textwidth]{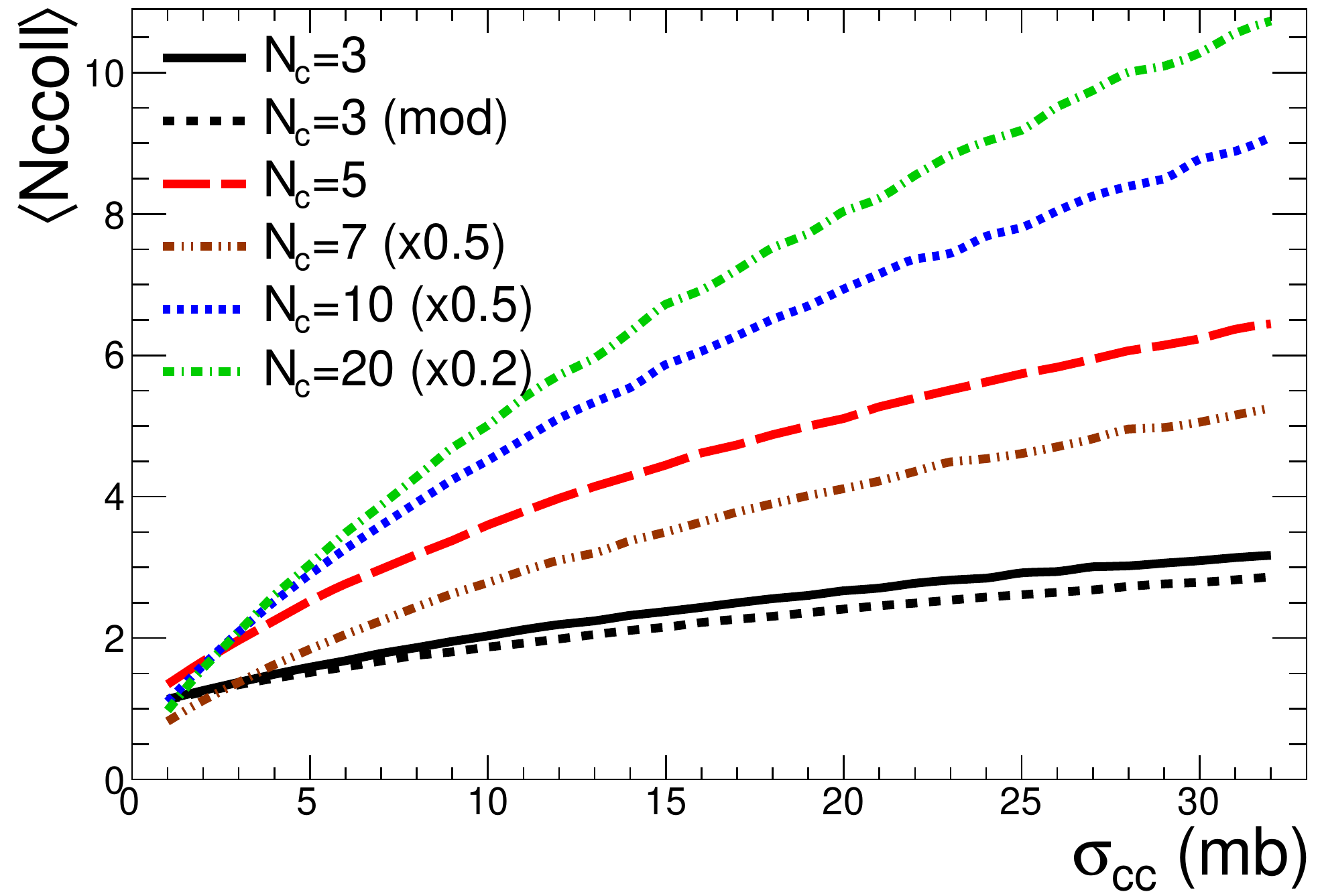}\hspace{0.125cm}
   \includegraphics[width=0.42\textwidth]{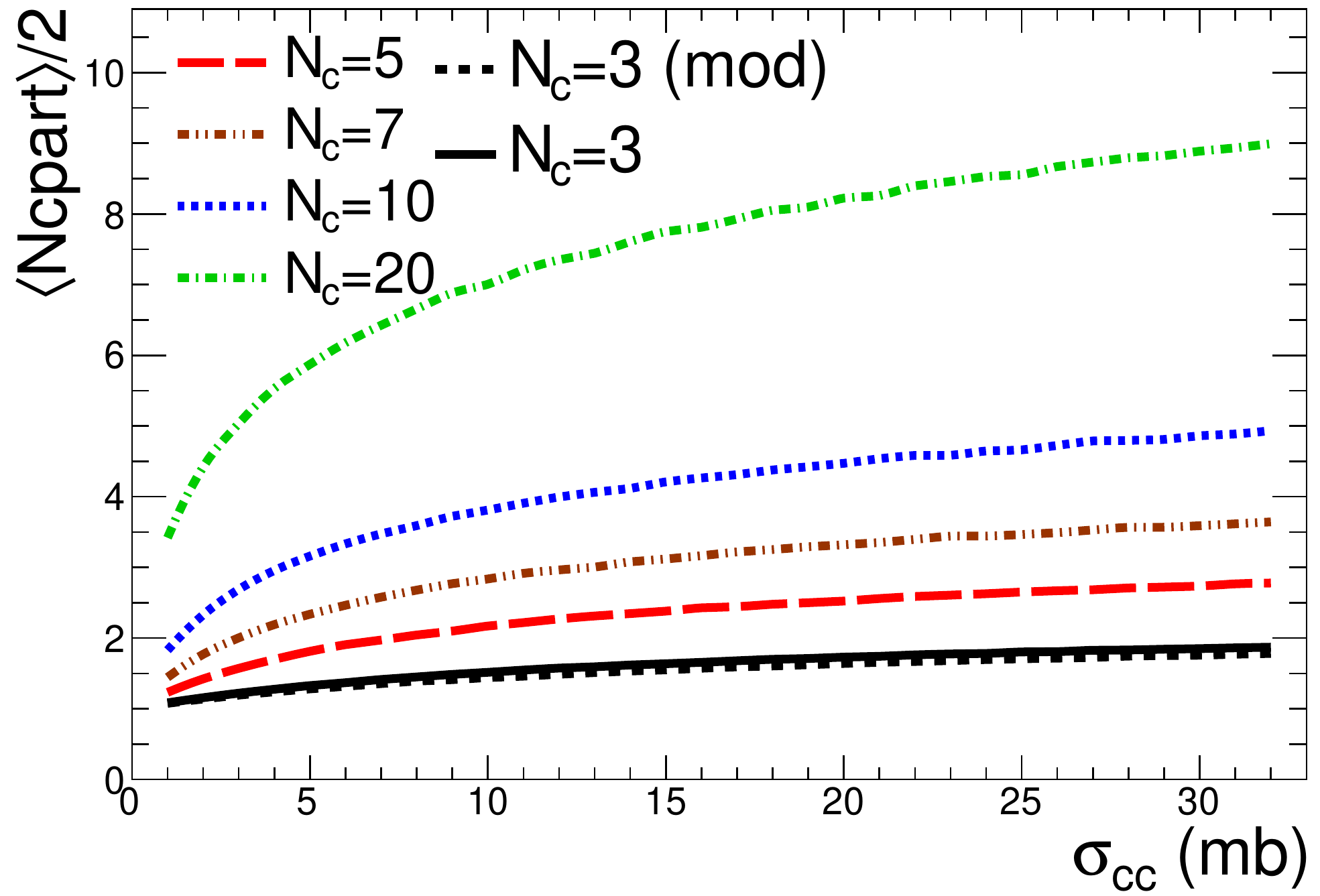}
   \includegraphics[width=0.42\textwidth]{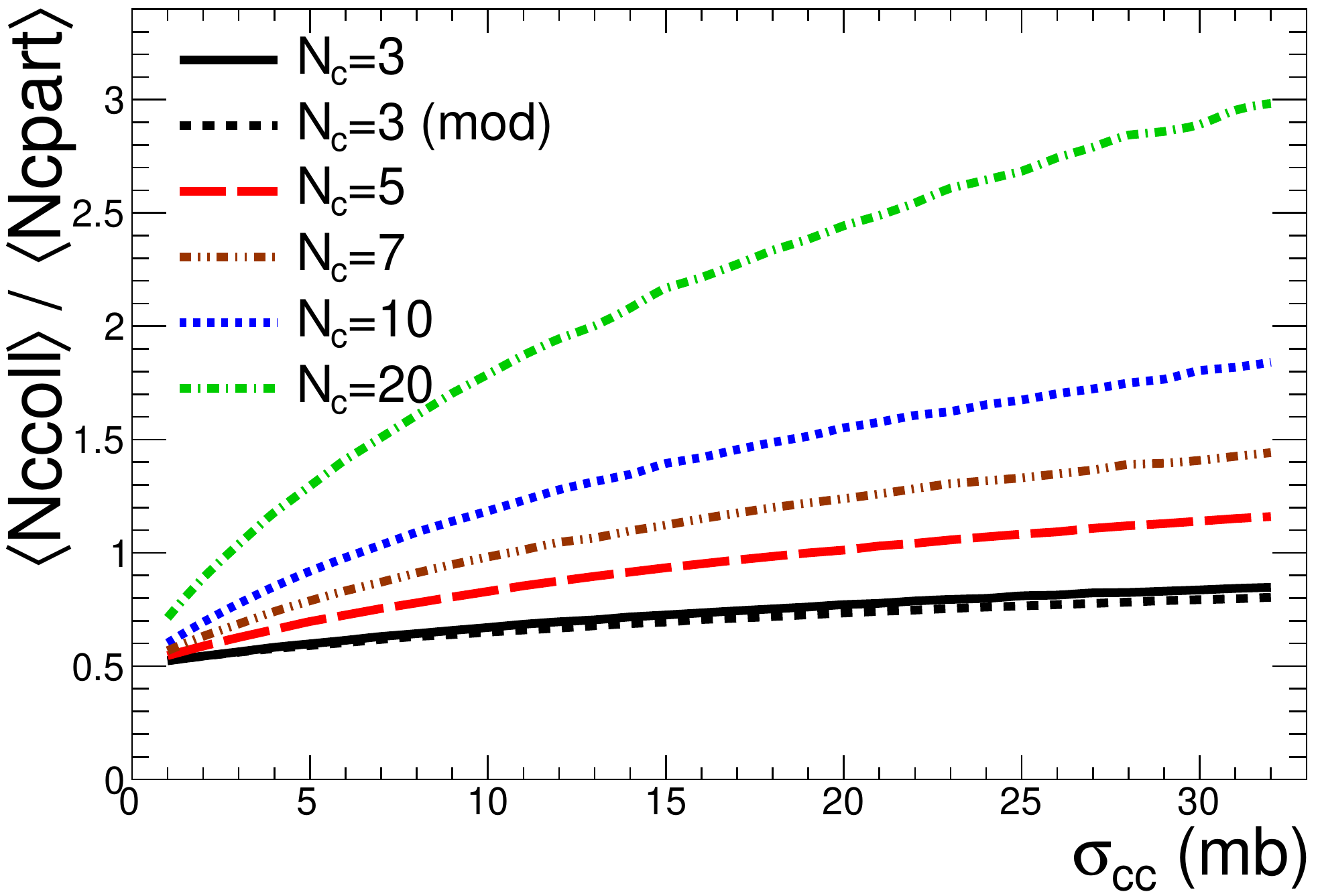}
   \caption{\label{fig:ppnccvsnpp}Average values of $\Nccoll$ (top) and $\Ncpart/2$ (middle), as well as of the ratio $\Nccoll/\Ncpart$ (bottom) versus $\sigcc$ for various $\Nc$ in pp collisions.}
\end{center}
\end{figure}
\begin{figure}[th]
\begin{center}
   \includegraphics[width=0.42\textwidth]{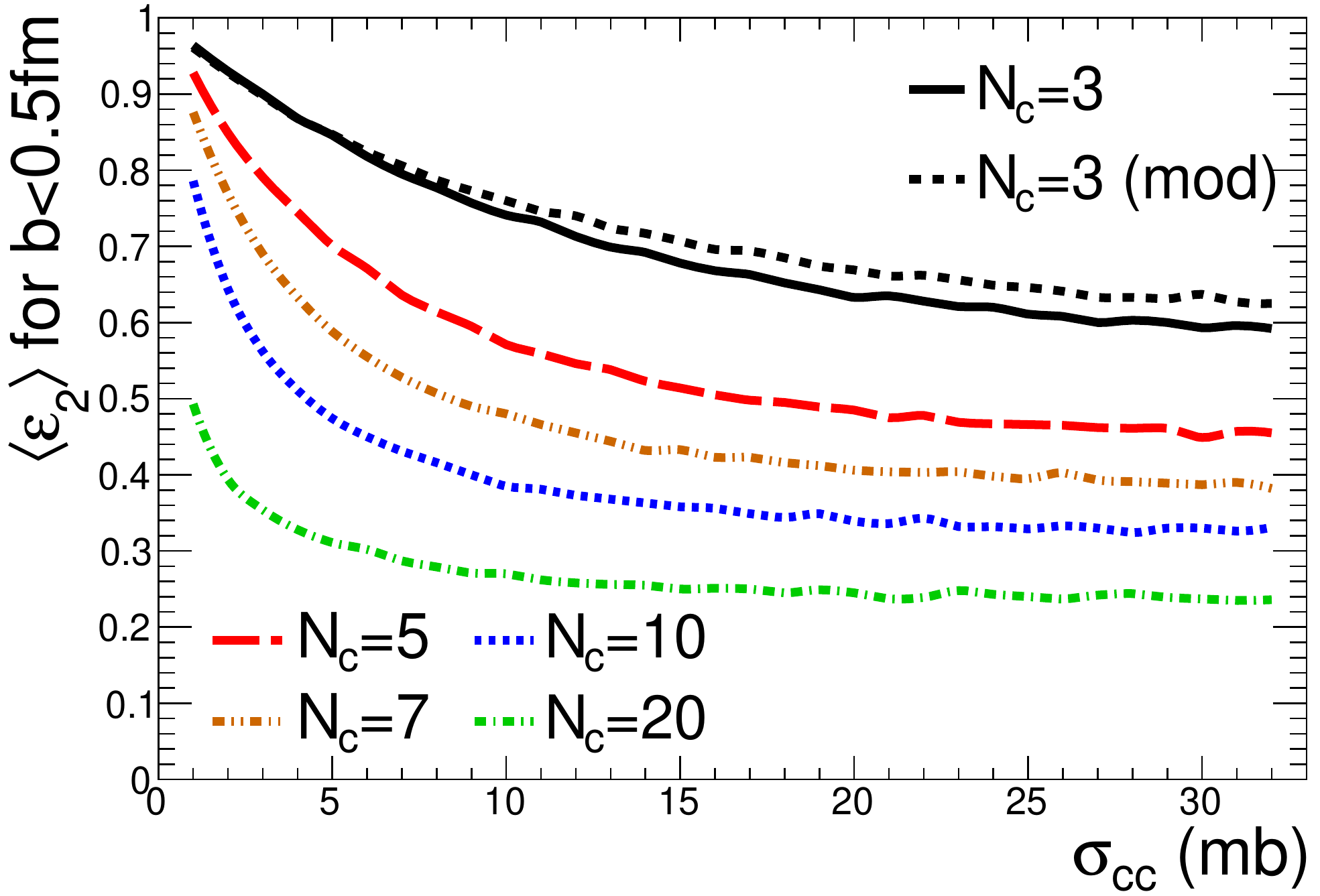}
   \includegraphics[width=0.42\textwidth]{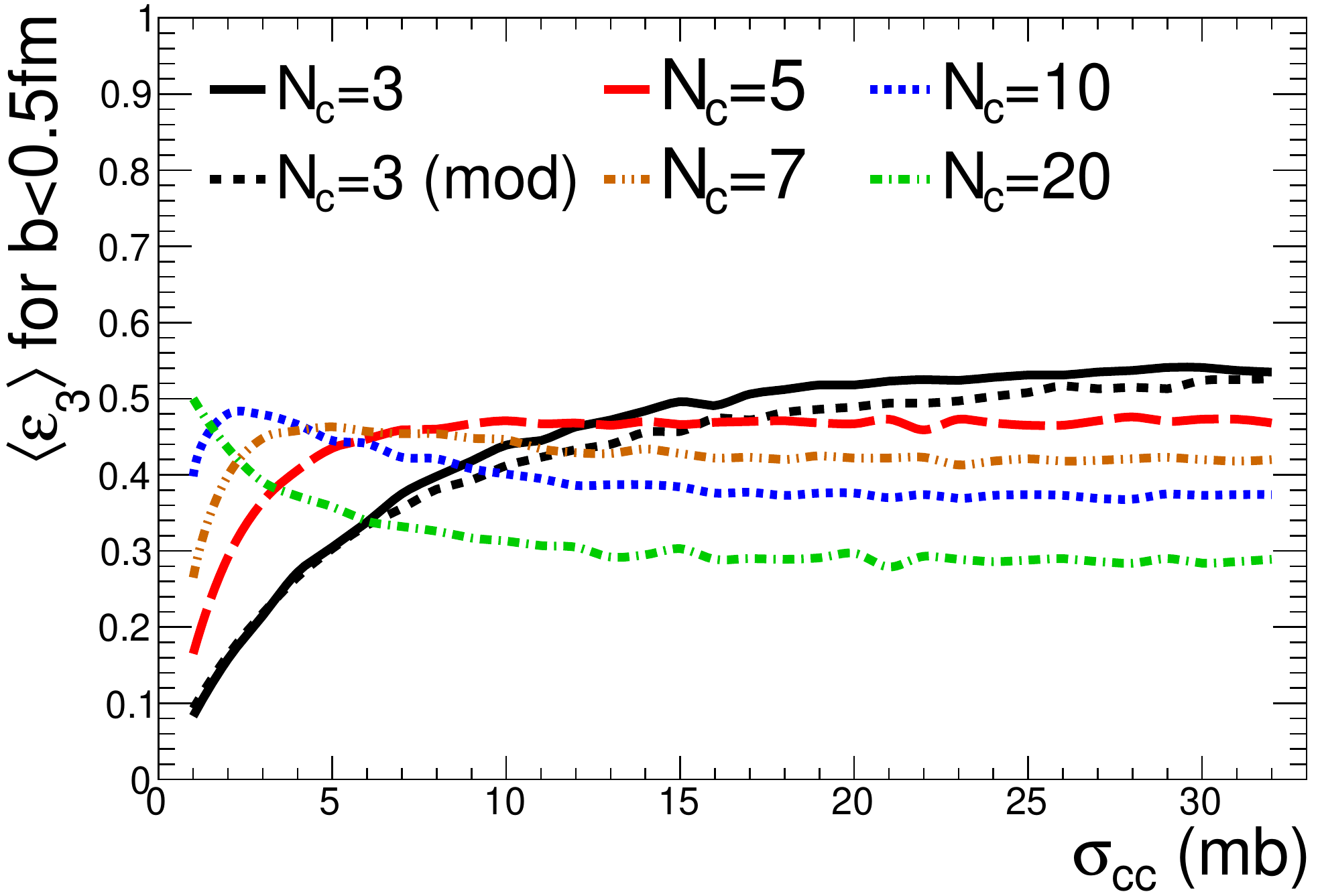}
   \caption{\label{fig:ppecc}Average eccentricity (top) and triangularity (bottom) for $b<0.5$~fm versus $\sigcc$ for various $\Nc$ in pp collisions.}
\end{center}
\end{figure}

\Figure{fig:xsecpbpb} shows the increase of the PbPb (top) and pPb (bottom) cross sections with $\sigcc$ for different values of $\Nc$ and the two ways to distribute the constituents, e.g.\ bound to nucleons or freely distributed inside the nucleus.
For the same parameters, the freely-distributing case always leads to a larger cross section than the bound case. 
In particular, for large $\sigcc$ and $\Nc$ the likelihood for peripheral collisions to occur increases significantly, making the total cross section exceed the value expected from geometrical considerations~(also visible in the right panel of \Fig{fig:impb}). 
For example, $8$~b corresponds to an effective radius of about $8$~fm (which is larger than $R+2a$ of Pb).

\section{Results}
\label{sec:results}
In this section, results of constituent Glauber model calculations are presented for pp and AA collision systems, for different input parameters.
\subsection{pp collisions}
\Figure{fig:ppnccvsnpp} shows average values of $\Nccoll$ and $\Ncpart/2$, as well as of the ratio $\Nccoll/\Ncpart$ versus $\sigcc$ for various $\Nc$ in pp collisions.
The resulting values increase with increasing $\Nc$ and $\sigcc$ compared to those at the nucleon level, which are $\Npart=2$, $\Ncoll=1$, and $\Ncoll/\Npart=0.5$.
For simplicity, $\av{\Nccoll}$ is also denoted as $\nu$ and $\av{\Ncpart}$ as $\mu$ in pp collisions.

\begin{figure}[th!]
\begin{center}
   \includegraphics[width=0.42\textwidth]{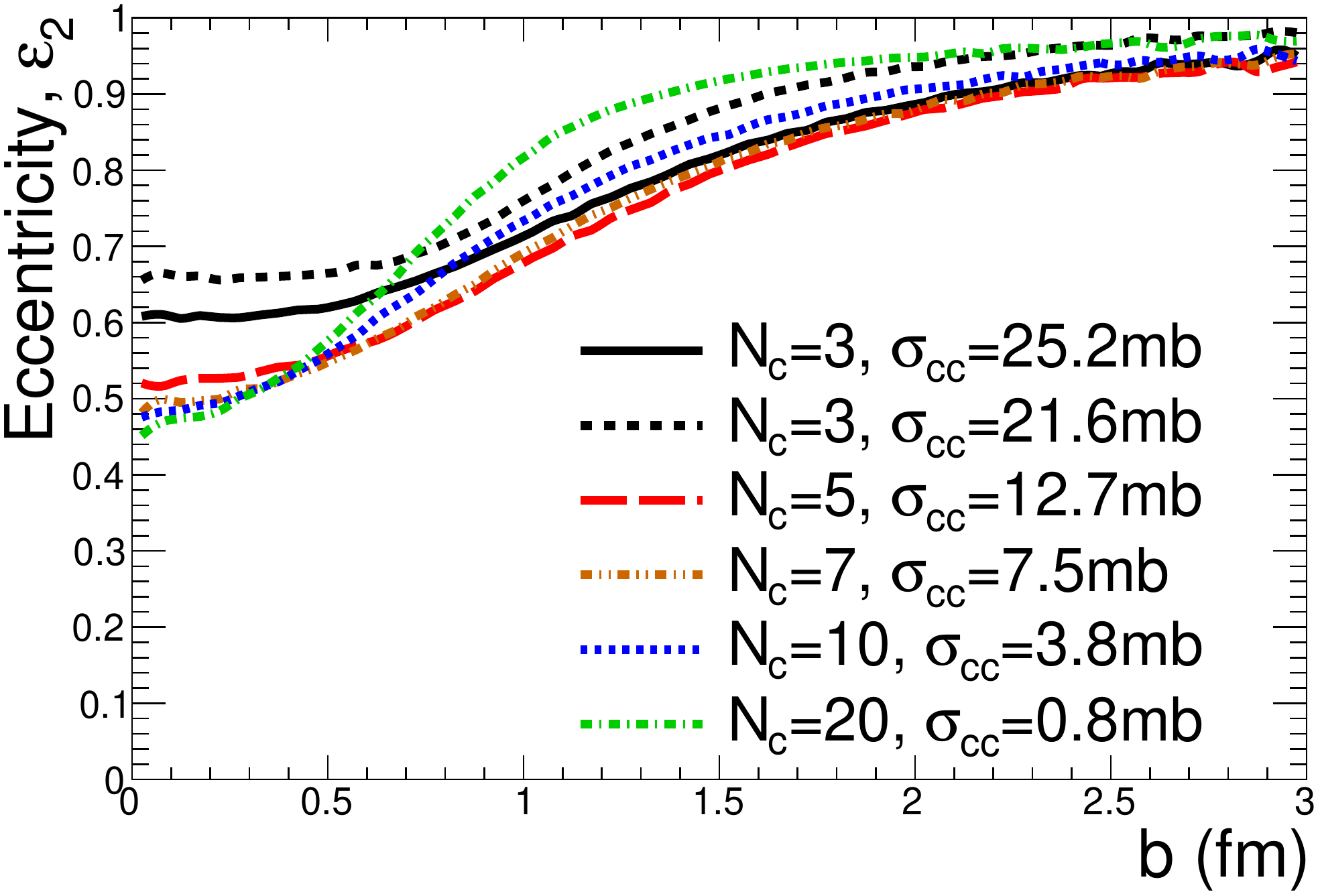}
   \includegraphics[width=0.42\textwidth]{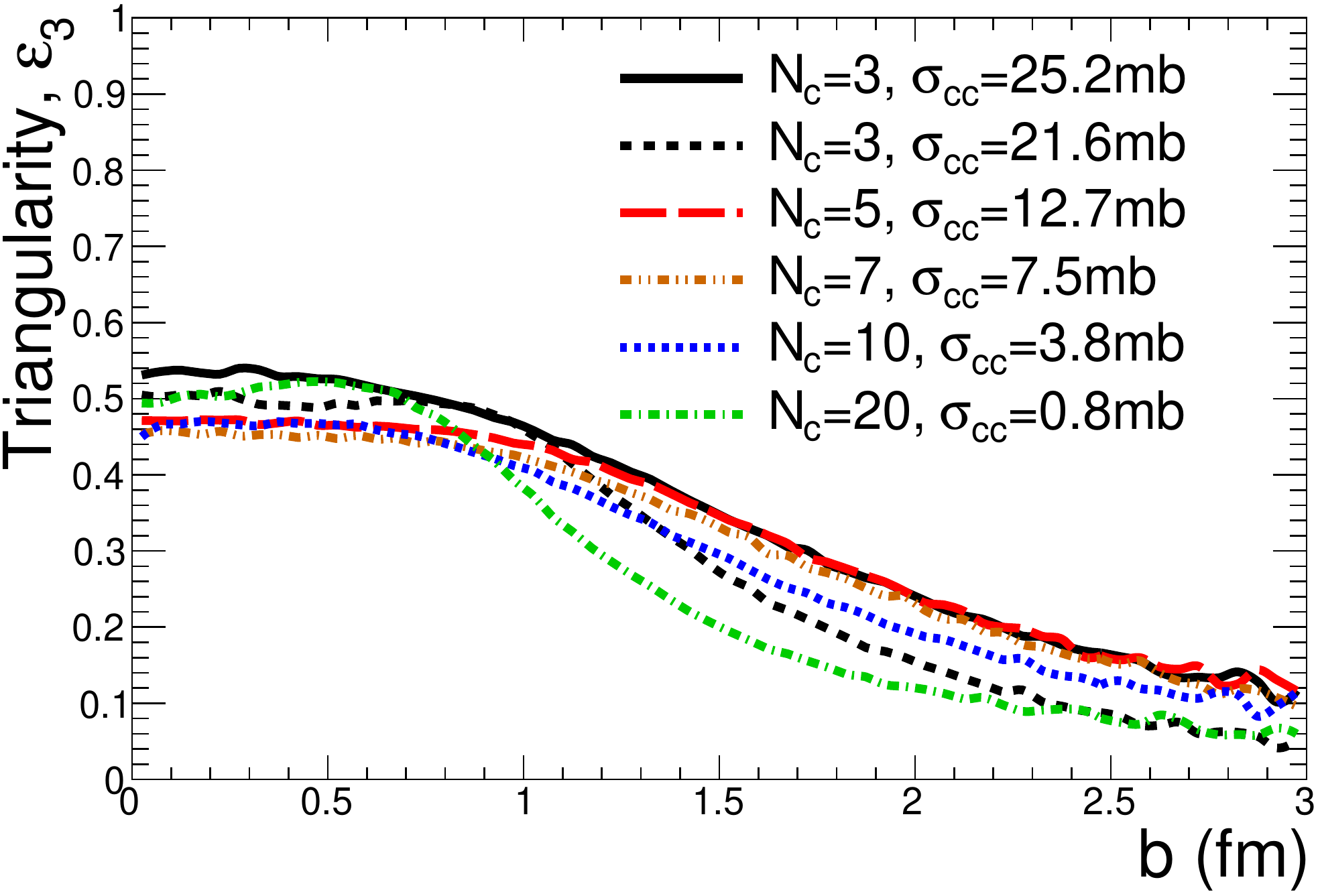}
   \caption{\label{fig:pp13ecc}Eccentricity (top) and triangularity (bottom) versus $b$ for various $\Nc$ in pp collisions at 13 TeV. The calculation for $\sigcc=21.60$~mb corresponds to the modified case.}
\end{center}
\end{figure}
\begin{figure}[th!]
\begin{center}
   \includegraphics[width=0.235\textwidth]{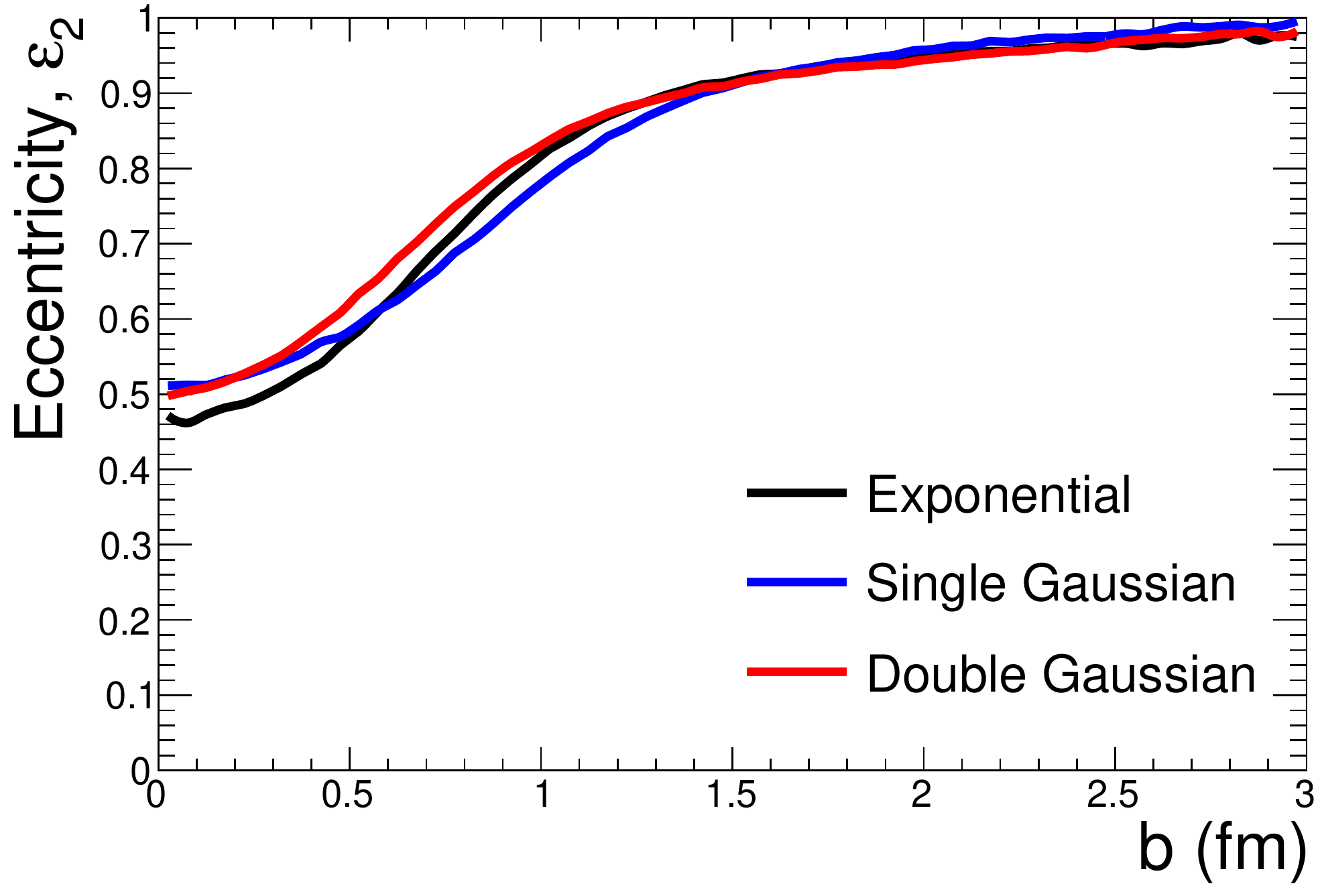}
   \includegraphics[width=0.235\textwidth]{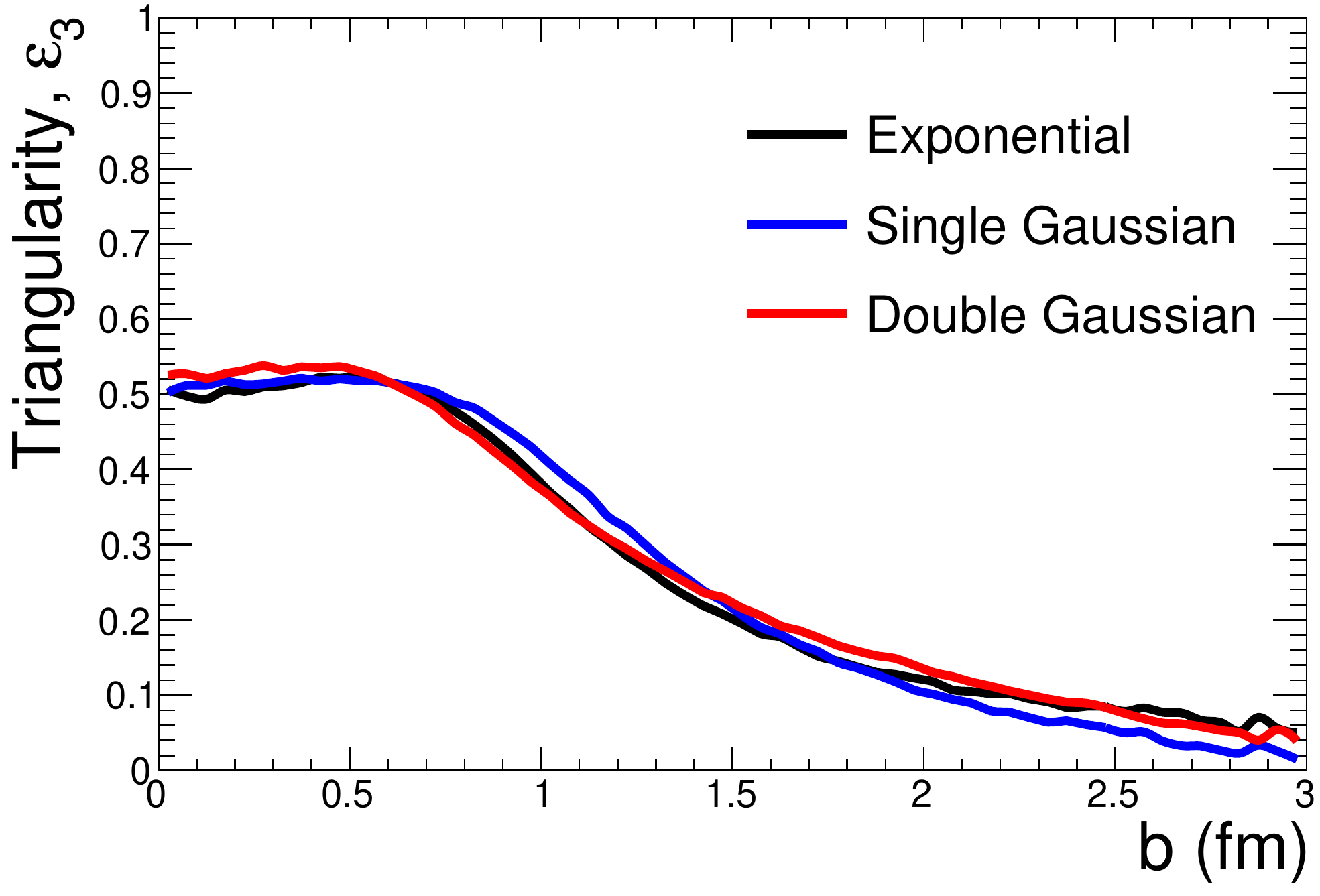}
   \caption{\label{fig:pp13eccprof}Eccentricity (left) and triangularity (right) versus $b$ for $\Nc=20$ and $\sigcc=0.8$ mb corresponding to pp collisions at 13 TeV using Exponential, Single and Double Gaussian density profiles~(see text). }
\end{center}
\end{figure}

\Figure{fig:ppecc} shows the dependence of the average eccentricity $\av{\varepsilon_{2}}$ and triangularity $\av{\varepsilon_{3}}$ versus $\sigcc$ for various $\Nc$ in central
pp collisions with $b<0.5$~fm. 
Increasing $\Nc$ and $\sigcc$ decreases the observed initial-state anisotropy as expected for a spherically symmetric system. 
In the limiting case, without sub-structure, $\varepsilon_{2}=1$ and $\varepsilon_{3}=0$.
\Figure{fig:pp13ecc} shows eccentricity and triangularity versus $b$ for a set of input parameters reflecting pp collisions at 13 TeV.
For central collisions ($b<0.5$~fm) $0.45<\varepsilon_{2}<0.65$ and $0.46<\varepsilon_{3}<0.54$ leading to scaled values of about $0.1$ and $0.02$ for measured values~\cite{Aad:2015gqa,Khachatryan:2016txc} of $v_2\sim0.05$ and $v_3\sim0.01$, respectively.
\Figure{fig:pp13eccprof} compares eccentricity and triangularity versus $b$ for $\Nc=20$ and $\sigcc=0.85$ mb corresponding to pp collisions at 13 TeV for different density profiles.
The first is the exponential (\Eq{eq:4}) profile, used so far.
The others are Single and Double Gaussian profiles, implemented in the impact-parameter dependent Glauber-like collision framework of PYTHIA8~\cite{Corke:2011yy}, and typically used to model multi-parton interactions.
The resulting distributions for $\varepsilon_{2}$ and $\varepsilon_{3}$ with the Single and Double Gaussian profiles do not differ from the standard case.

\begin{table}[t]
\begin{center}
  \begin{tabular}{c|c||c|c|c|c|c}
    $\Nc$ & $\sigcc$~(mb) & $\mu$ & $\nu$ & $\signn$~(mb) & $\sigpbpb$ (b) & $\sigpbpb^{\rm free}$ (b)\\
    \hline
    \hline
     3 & 21.1   & 3.5 & 2.7 & 70.0 & 7.94 & 8.24 \\ 
     3$^{*}$ & 17.9    & 3.2 & 2.3 & 70.1 & 7.94   \\ 
     5 & 10.3   & 4.4 & 3.7 & 70.1 & 7.94 & 8.46 \\
     7 & 5.7  & 4.9 & 4.0 & 70.0 & 7.93 & 8.56 \\
    10 & 2.8  & 5.2 & 4.0 & 70.0 & 7.94 & 8.62 \\
    \hline
     3 & 14.4 & 3.3 & 2.3 & 55.4 & 7.74 \\
     3$^{*}$ & 11.9 & 3.0 & 2.0 & 53.6 & 7.71\\
     3 & 8.8  & 3.0 & 1.9 & 40.9 & & 7.74 \\
     5 & 6.4  & 3.9 & 2.8 & 55.8 & 7.75 \\
     5 & 2.9  & 3.1 & 1.9 & 37.2 & & 7.76 \\
    10 & 1.7  & 4.4 & 2.9 & 56.8 & 7.78 \\
  \end{tabular}
  \caption{\label{tab:pbpbvals}Values for PbPb collisions at $\signn=5.02$ TeV. Input parameters are $\Nc$, $\sigcc$ and the way the constituents are distributed (bound, modified and free). Output values are $\mu=\av{\Ncpart}$, $\nu=\av{\Nccoll}$, $\signn$ and $\sigpbpb$ as well as $\sigpbpb^{\rm free}$ for the freely distributing case. The modified cases are indicated with $^{*}$. The parameters in rows above the horizontal line are chosen to match $\signn=70$~mb, while those below the horizontal line $\sigpbpb=7.7$~b. 
}
\end{center}  
\end{table}
\begin{table}[t]
\begin{center}
  \begin{tabular}{c|c||c|c|c|c|c}
    $\Nc$ & $\sigcc$~(mb) & $\mu$ & $\nu$ & $\signn$~(mb) & $\sigauau$ (b) & $\sigauau^{\rm free}$ (b)\\
    \hline
    \hline
     3      & 6.3  & 2.8 & 1.7 & 33.0 & 6.89 & 7.03 \\
     3$^{*}$ & 5.8  & 2.6 & 1.6 & 33.0 & 6.91 \\ 
     5      & 2.4  & 3.0 & 1.8 & 33.1 & 7.01 & 7.16\\
    \hline
     3 & 3.6 & 2.5 & 1.4 & 22.4 & 6.65 \\
     3$^{*}$ & 3.8 & 2.5 & 1.4 & 24.1 & 6.67\\
     3 & 3.2 & 2.5 & 1.4 & 20.5 & & 6.66 \\
     5 & 1.4 & 2.6 & 1.5 & 23.2 & 6.70 \\ 
  \end{tabular}
  \caption{\label{tab:auauvals}Values for AuAu collisions at $\signn=19.6$ TeV, with $\signn=33$~mb and $\sigauau=6.7$~b. See description in \Tab{tab:pbpbvals} for more information.}
\end{center}  
\end{table}

\subsection{AA collisions}
The results for AA collisions\co{, discussed in this section,} are presented for PbPb collisions at $\snn=5.02$ TeV, and AuAu collisions at $\snn=19.6$ GeV, respectively.
The calculations are done for various choices of $\Nc$ and $\sigcc$, as well as various ways to distribute the constituents, i.e.\ bound, modified and free cases. 
The parameters, which are summarized in \Tab{tab:pbpbvals} and \Tab{tab:auauvals}, have been set to either match the corresponding $\signn$ or $\sigpbpb$.
When fixing $\signn$, the calculated cross sections exceed $\sigpbpb$ by about 3\% in the constrained and by up to 12\% in the free case. 
When fixing $\sigpbpb$, the effective $\signn$ are lower by up to 25\% in the constrained and up to 50\% in the free case.
The idea is to compare the results for a set of parameters that lead to the similar measurable quantities~(the uncertainty of the measured cross sections is on the level of $5$--$10$\%) 
to study the robustness of conclusions with respect to a priori unfalsifiable assumptions.

\begin{figure}[t!]
\begin{center}
   \includegraphics[width=0.235\textwidth]{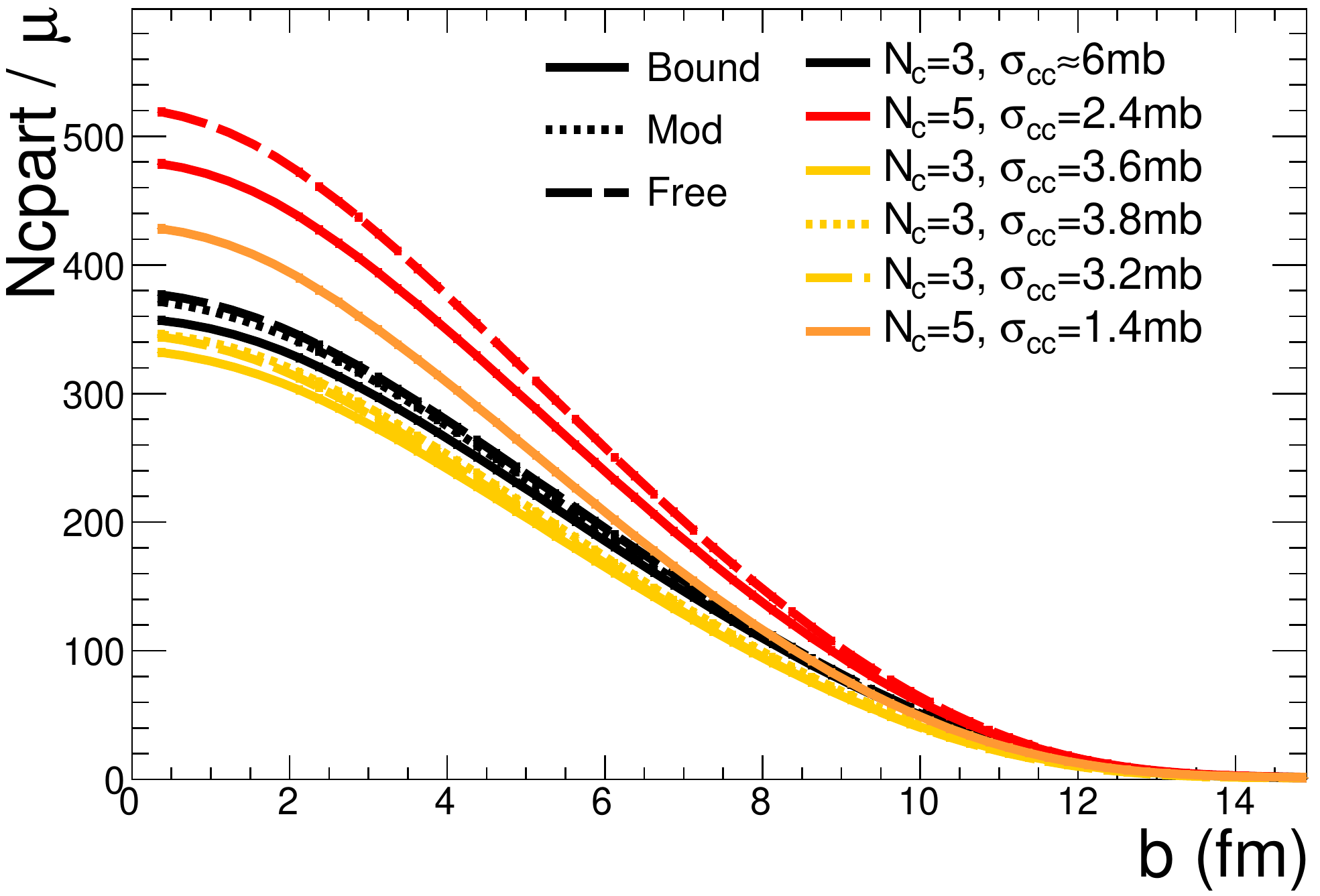}
   \includegraphics[width=0.235\textwidth]{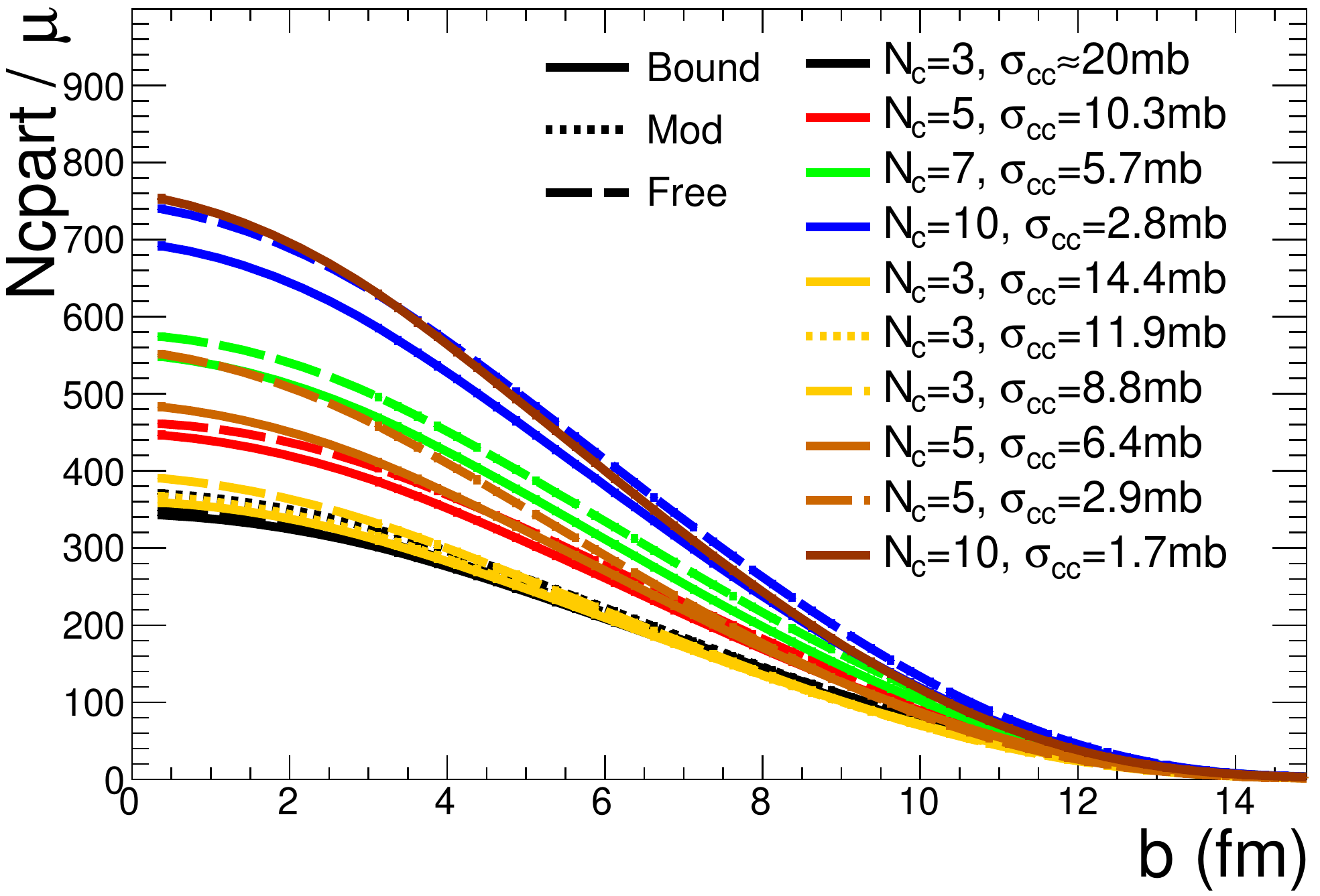}
   \includegraphics[width=0.235\textwidth]{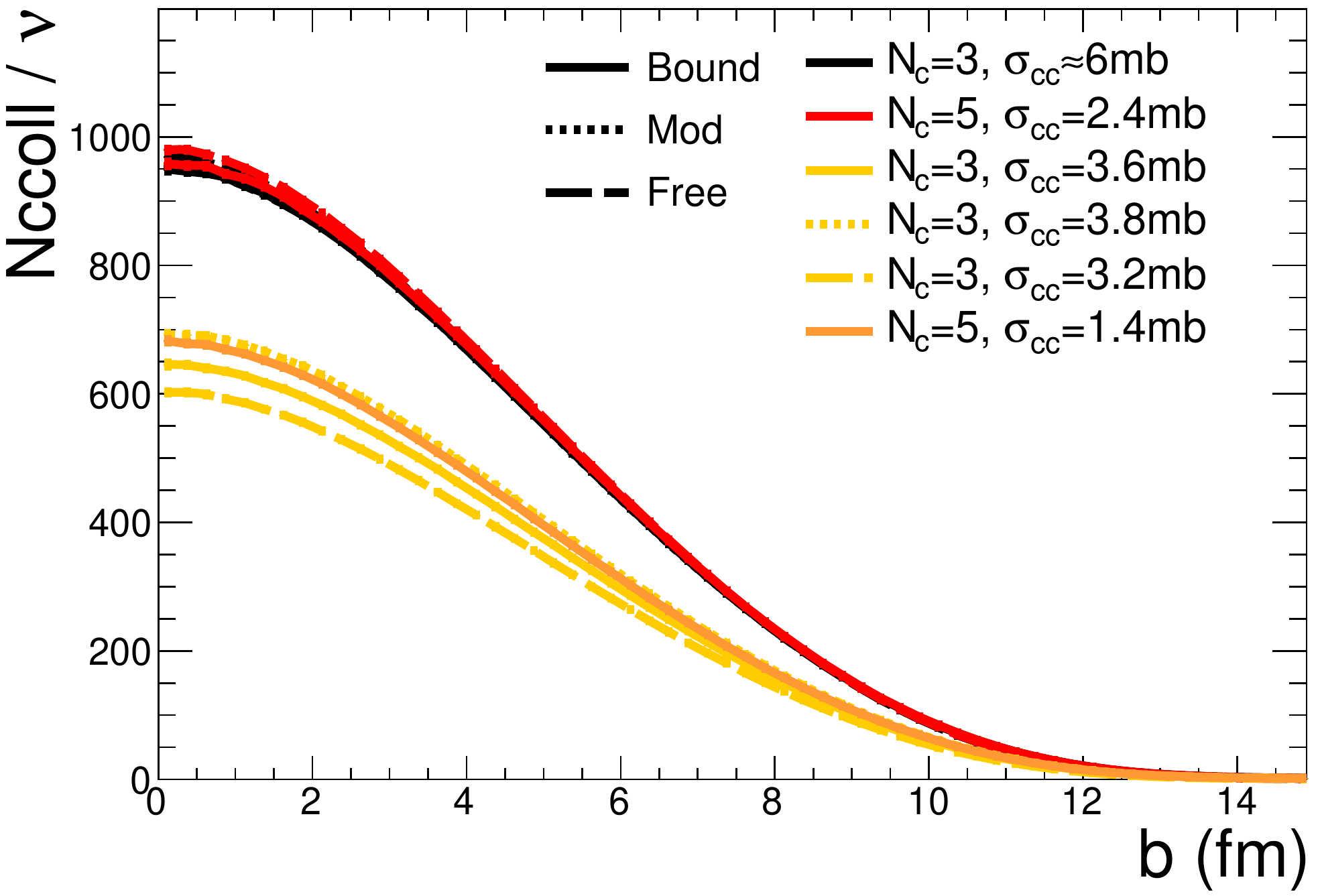}
   \includegraphics[width=0.235\textwidth]{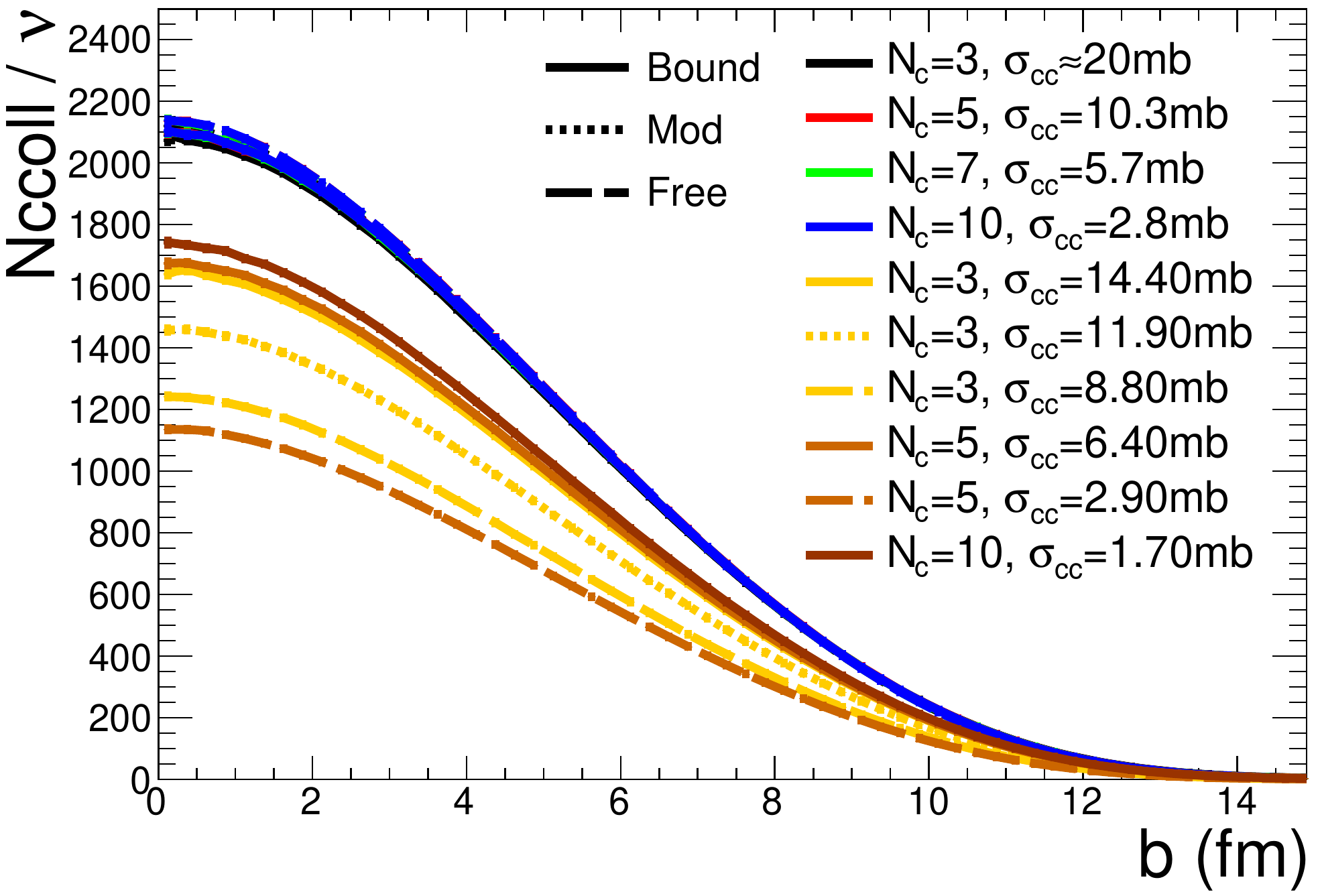}
   \caption{\label{fig:ncvsb}$\Ncpart/\mu$ (top) and $\Nccoll/\nu$ (bottom panels) for AuAu (left) and PbPb (right panels) collisions. The parameters for the calculations are summarized in \Tab{tab:pbpbvals} and \Tab{tab:auauvals}.}
\end{center}
\end{figure}
\begin{figure}[th!]
\begin{center}
   \includegraphics[width=0.45\textwidth]{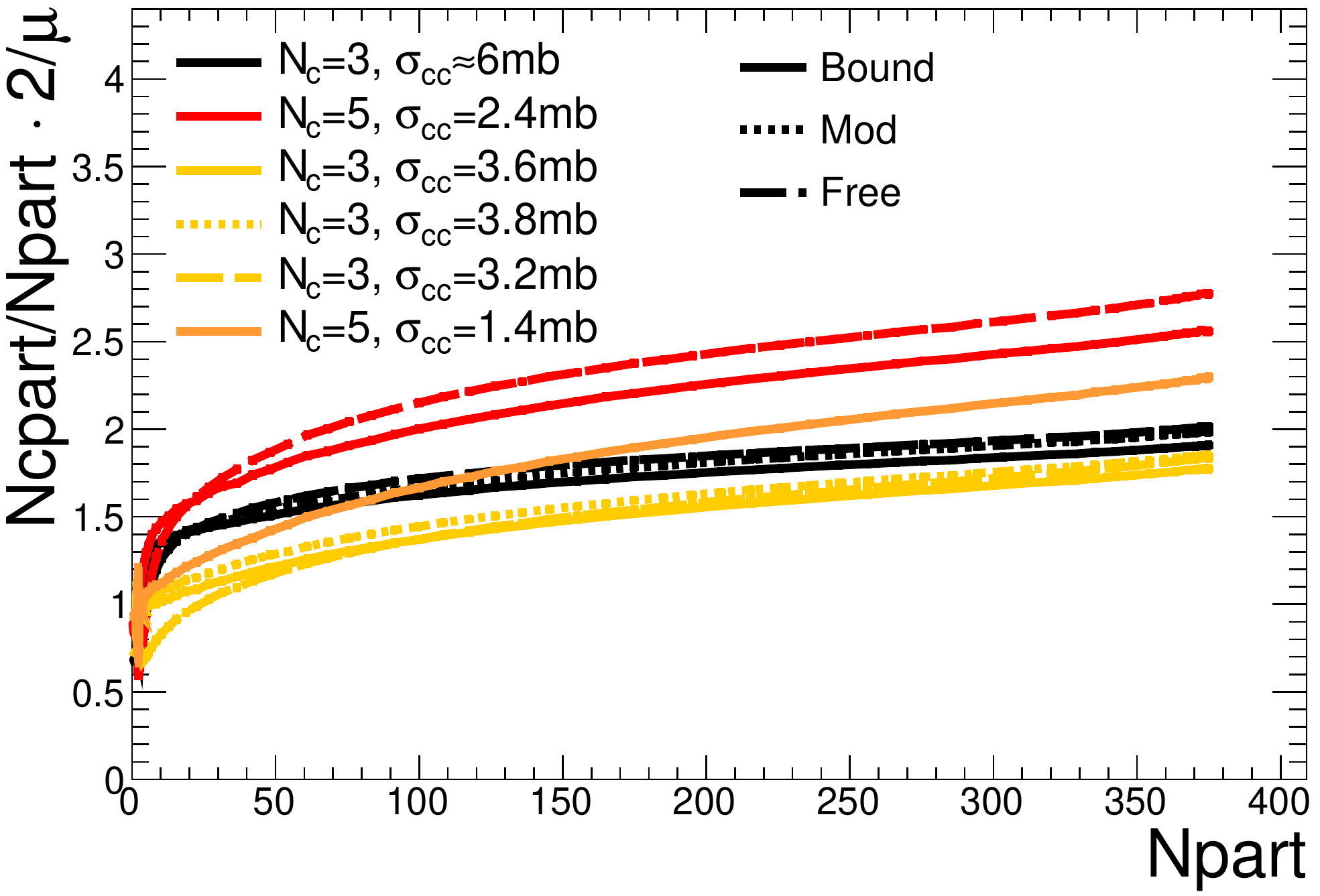}
   \includegraphics[width=0.45\textwidth]{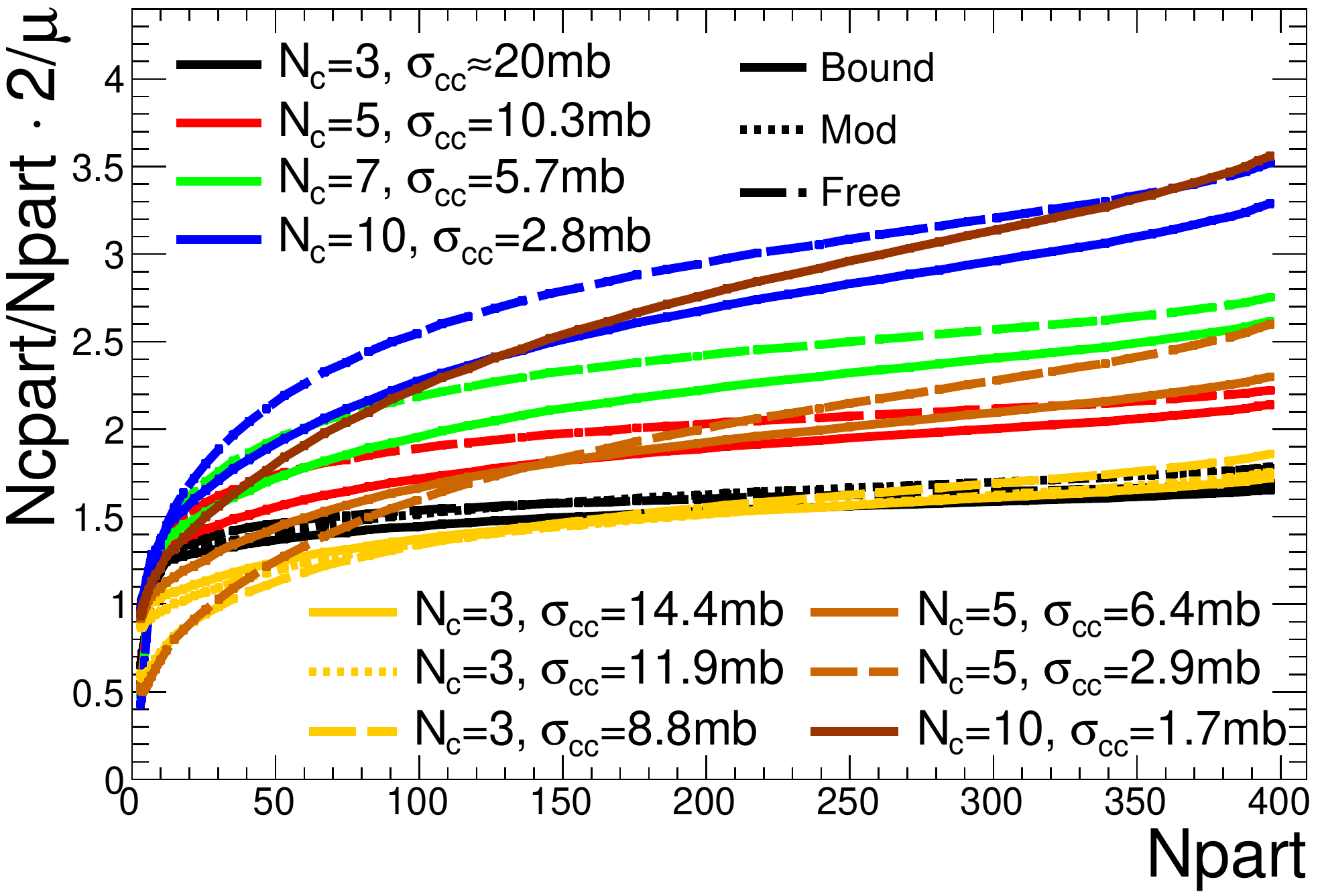}
   \caption{\label{fig:ncnormvsnpart}Ratio $\Ncpart/\Npart$ normalized to $\mu/2$ for AuAu (top) and PbPb (bottom panel) collisions. The parameters for the calculations are summarized in \Tab{tab:pbpbvals} and \Tab{tab:auauvals}.}
\end{center}
\end{figure}
\begin{figure}[th!]
\begin{center}
   \includegraphics[width=0.45\textwidth]{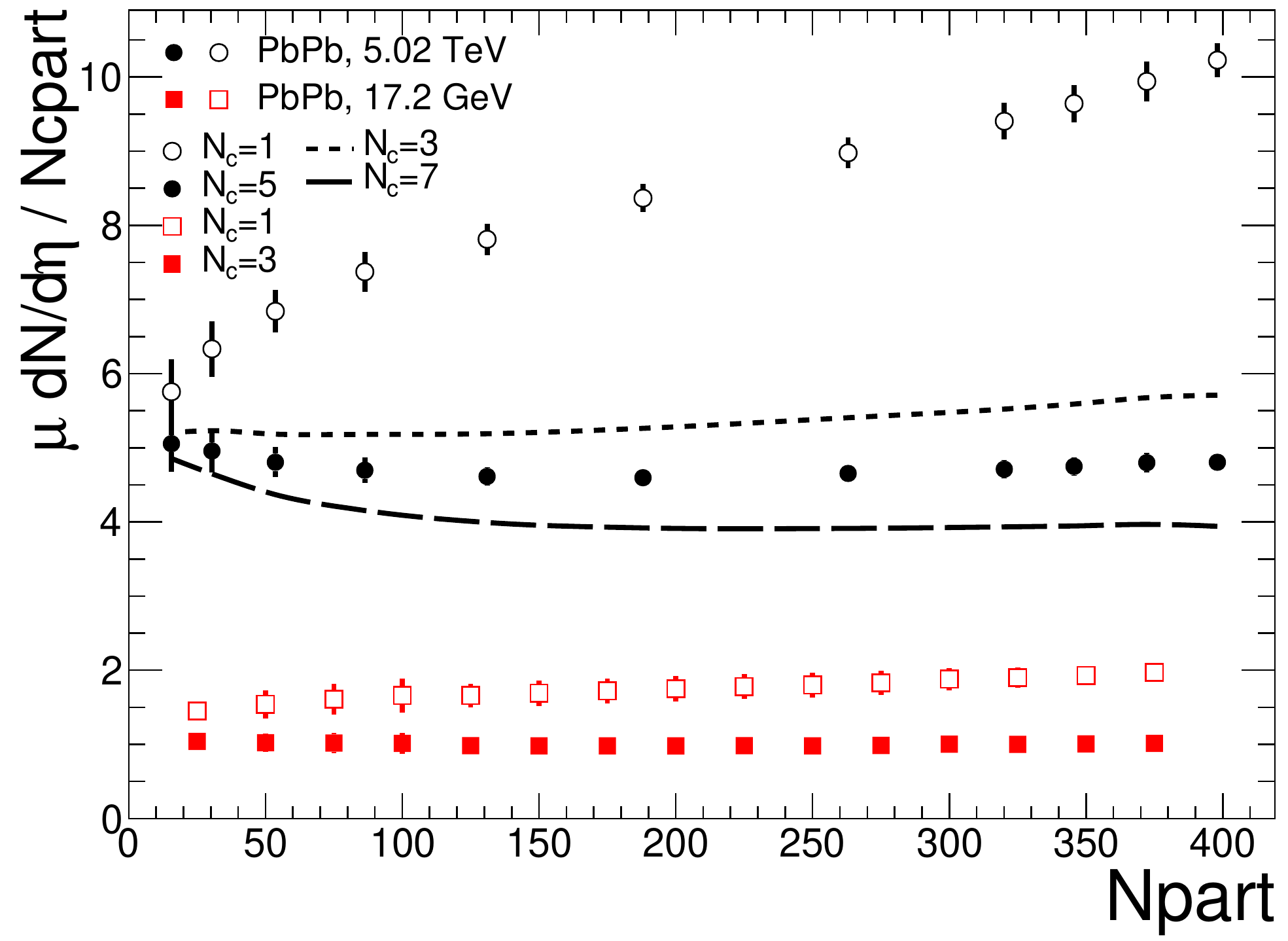}
   \caption{\label{fig:dndeta}Values of ${\rm d}N/{\rm d}\eta$ in PbPb collisions at $\snn=17.2$ GeV and $\snn=5.02$ TeV scaled by $\Ncpart/\mu$ for sub-nucleon~($\Nc>1$) and $\Npart/2$ for nucleon~($\Nc=1$) participants. The data are from~\cite{Adler:2004zn,Adam:2015ptt}, drawn with only point-to-point uncorrelated systematic uncertainties. The 17.2 GeV data are scaled using $\Nc=3$~($\sigcc=5.5$~mb, modified case). The 5.02 TeV data are scaled using $\Nc=5$~($\sigcc=10.3$~mb). The lines show the central points if the data were scaled by $\Nc=3$~($\sigcc=17.9$~mb, modified case) and $\Nc=7$~($\sigcc=5.7$~mb), respectively.}\vspace{-0.8cm} 
\end{center}
\end{figure}

\Figure{fig:ncvsb} shows $\Ncpart/\mu$ and $\Nccoll/\nu$ versus $b$, where $\mu=\av{\Ncpart}$ and $\nu=\av{\Nccoll}$ in pp collisions, respectively. 
\Figure{fig:ncnormvsnpart} shows the ratio $\Ncpart/\Npart$ versus $\Npart$ normalized to $\mu/2$.
The calculations are performed in small bins of $b$ and then matched to the corresponding $\Npart$ at the nucleon level, since in peripheral events $\av{\Npart}$ for events selected with $\Ncpart>0$ slightly differs from those calculated at the nucleon level and selected with $\Npart>0$.
In particular for $\Nc\le5$, the shape is similar to that of the measured $2{\rm d}N/{\rm d}\eta/\Npart$~\cite{Alver:2010ck,Aamodt:2010cz,Adam:2015ptt}.
The inverse of what is plotted, i.e.\ $\mu/2\Npart/\Ncpart$, would be the factor needed to translate the measurements scaled by $\Npart/2$ to $\Ncpart/\mu$.
Hence, the correction would affect the shape the strongest for peripheral events, making $\mu{\rm d}N/{\rm d}\eta/\Ncpart$ approximately flat. 

\begin{figure}[th!]
\begin{center}
   \includegraphics[width=0.45\textwidth]{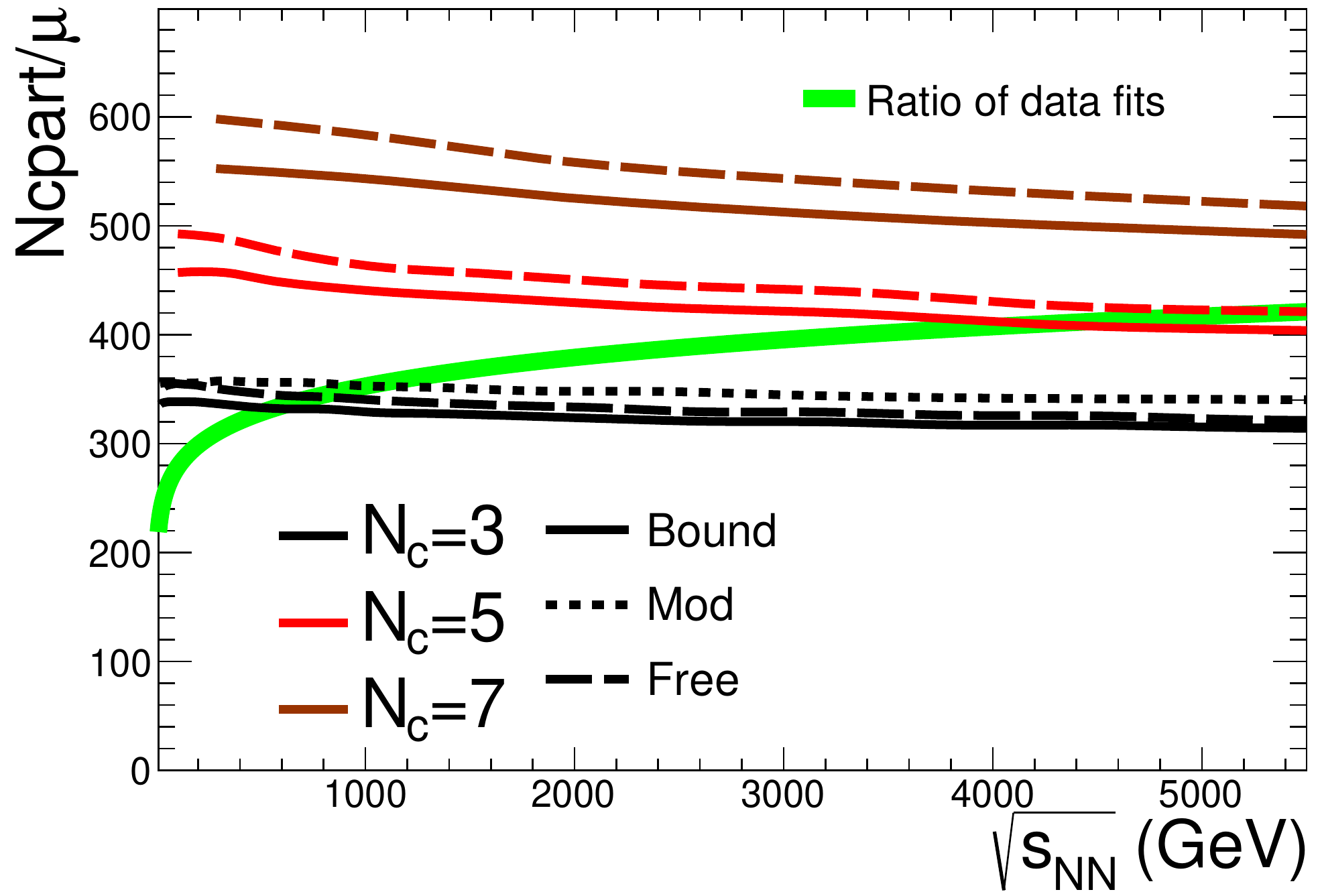}
   \caption{\label{fig:ratio}Ratio of fits to central AA (scaled by 160 to approximately account for $\Npart/2$) and inelastic pp collisions compared to $\Ncpart/\mu$ from constituent Glauber calculations for $b<3.5$~fm. The values for the power law fits are taken from~\cite{Adam:2015ptt}.}
\end{center}
\end{figure}
\begin{figure}[th!]
\begin{center}
   \includegraphics[width=0.45\textwidth]{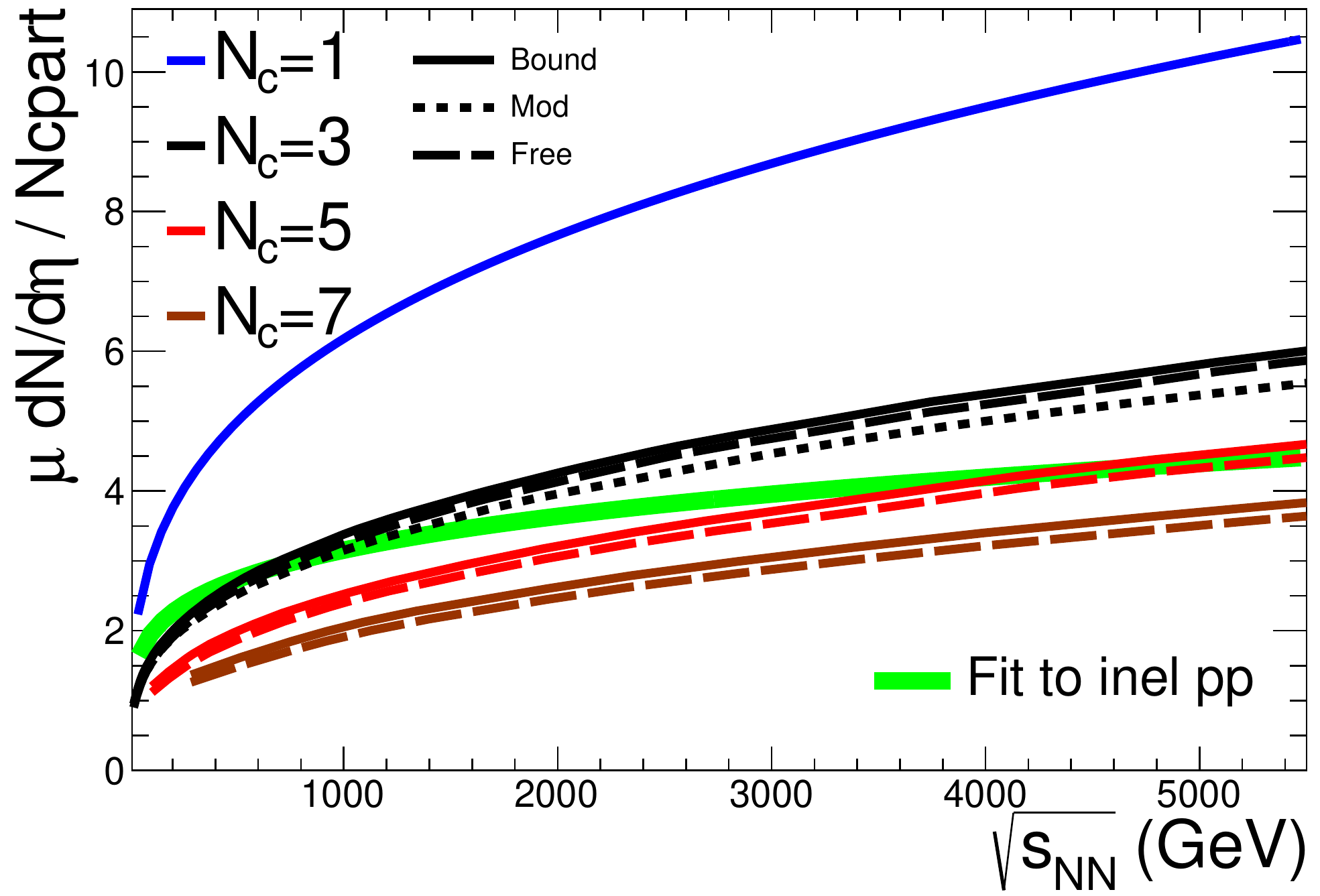}
   \caption{\label{fig:abs}Power-law fit of ${\rm d}N/{\rm d}\eta$ from inelastic pp collisions compared to scaled central AA data. The AA curves are obtained from a power-law fit to $2{\rm d}N/{\rm d}\eta/\Npart$ scaled by $\mu/\Ncpart$ (multiplied by 160 to approximately account for $\Npart/2$) for constituent Glauber calculations with $b<3.5$~fm. In the case of $\Nc=1$, $\Ncpart=\Npart$ and $\mu=2$, the shown curve essentially represents the original fit to $2{\rm d}N/{\rm d}\eta/\Npart$. The values for the power law fits are taken from~\cite{Adam:2015ptt}.}
\end{center}
\end{figure}

Indeed, this is directly demonstrated in \Fig{fig:dndeta}, which shows ${\rm d}N/{\rm d}\eta$ in PbPb collisions at $\snn=17.2$ GeV~\cite{Adler:2004zn} and $\snn=5.02$ TeV~\cite{Adam:2015ptt} scaled by $\Ncpart/\mu$ for sub-nucleon~($\Nc>1$) and $\Npart/2$ for nucleon~($\Nc=1$) participants.\footnote{The data at $\snn=17.2$ GeV are used as proxy for the AuAu at $\snn=19.6$ GeV, since they were measured over a larger range in centrality. The corresponding $\signn$ is $32$~mb.}
Using $\Nc=3$ and $5$ for $\snn=17.2$~GeV and $5.02$ TeV, respectively, approximately flatten the scaled data, which when fit with a first order polynomial exhibit a slope consistent with zero~($0.0002\pm0.0004$ and $0.0000\pm0.0002$, respectively).
For the 5.02 TeV data also the cases $\Nc=3$ and $7$ are shown, which exhibit a small positive~($0.0016\pm0.0004$) and negative~($-0.0009\pm0.0003$) slope, respectively.
This may be an indication that the effective partonic degrees of freedom relevant for soft particle production are on average about $5$ at high energy, and about $3$ at lower collision energy.

This is further investigated by comparing particle production in central AA to pp collision data.
The mid-rapidity ${\rm d}N/{\rm d}\eta$ in central AA collisions scaled by $\Npart$ compared to that in inelastic pp collisions turned out to rise stronger with collision energy, with $s_{\rm NN}^{0.155}$ rather than $s_{\rm NN}^{0.103}$, respectively~\cite{Adam:2015ptt}.
To evaluate if normalizing by constituent instead of nucleon participants would lead to a more similar behavior, the ratio $\Ncpart/\mu$ has been computed for various $\Nc$ and several ways to distribute the sub-nuclear degrees of freedom.
The computed ratios for $b<3.5$~fm are shown versus $\snn$ in \Fig{fig:ratio}, and found to slightly decrease with increasing $\snn$.
This trend can be compared to data, using the ratio of the power-law fits to the central AA and the inelastic pp data taken from \Ref{Adam:2015ptt}.
The ratio of the power-law fits is scaled by $160$ to roughly account for normalizing the central AA data by $\Npart/2$, since for central AuAu at $\snn=19.6$~GeV $\Npart\approx340$, while $\Npart\approx385$ for PbPb at $\snn=5.02$~TeV.
As can be seen in the figure, the data exhibit a different trend, i.e.\ the ratio is slightly rising with $\snn$.
The comparison between data and calculations does not reveal a preferred constant value for $\Nc$.
Instead, at lower energy $\Nc=3$, while at higher energy $\Nc=5$ is supported by the data, indicating that the number of relevant partonic degrees of freedom increases with increasing collision energy.
On an absolute scale, \Figure{fig:abs} indeed confirms that scaling with $\Ncpart/\mu$ for $\Nc=3$ or $5$ leads to a more similar collision energy dependence of central AA and inelastic pp data than based on $\Npart$ (labeled with $\Nc=1$).
In particular, it is important to realize that while the collision energy varies by three orders of magnitude, the scaled ${\rm d}N/{\rm d}\eta$ only changes by a factor $2$.

\begin{figure}[t]
\begin{center}
   \includegraphics[width=0.235\textwidth]{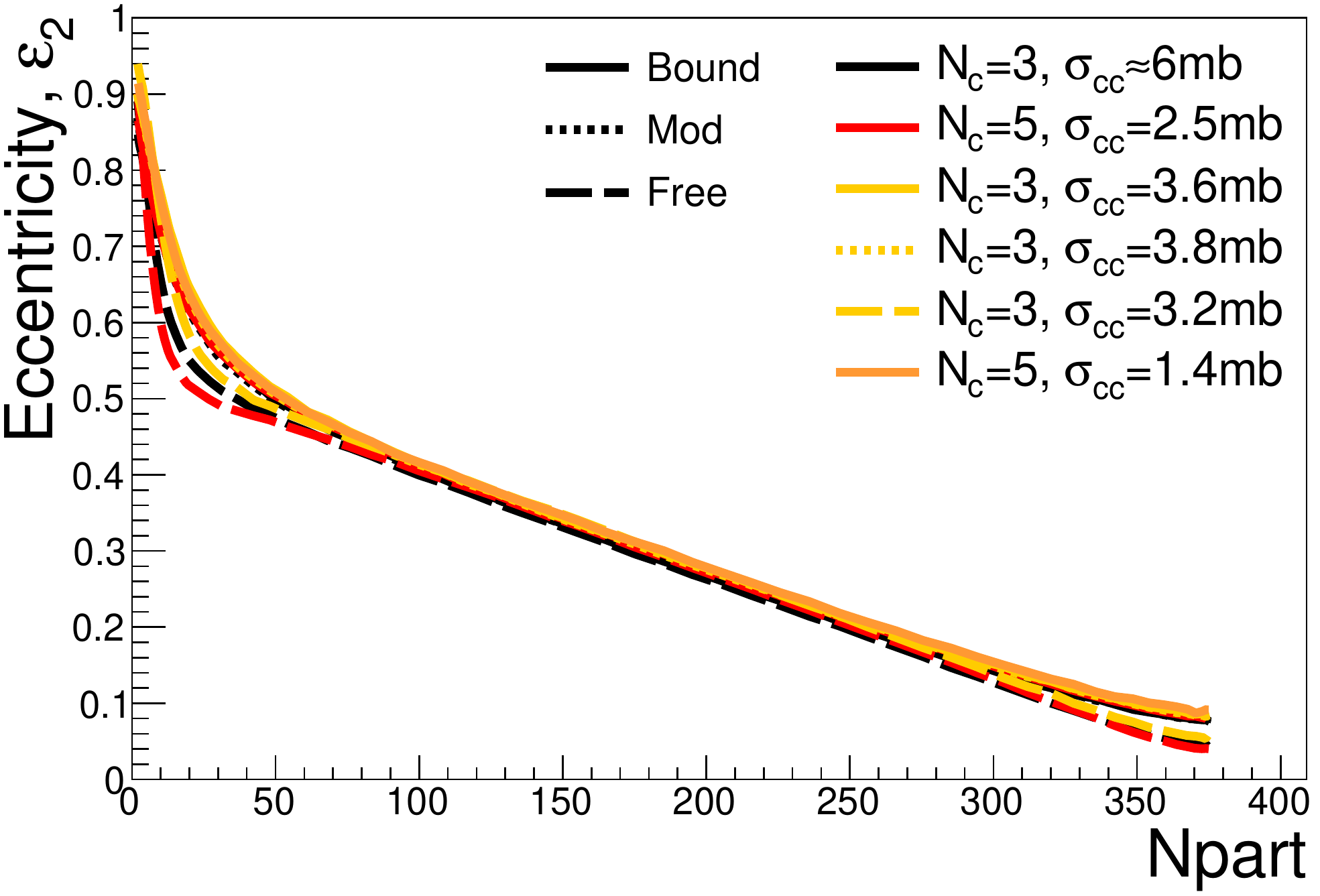}
   \includegraphics[width=0.235\textwidth]{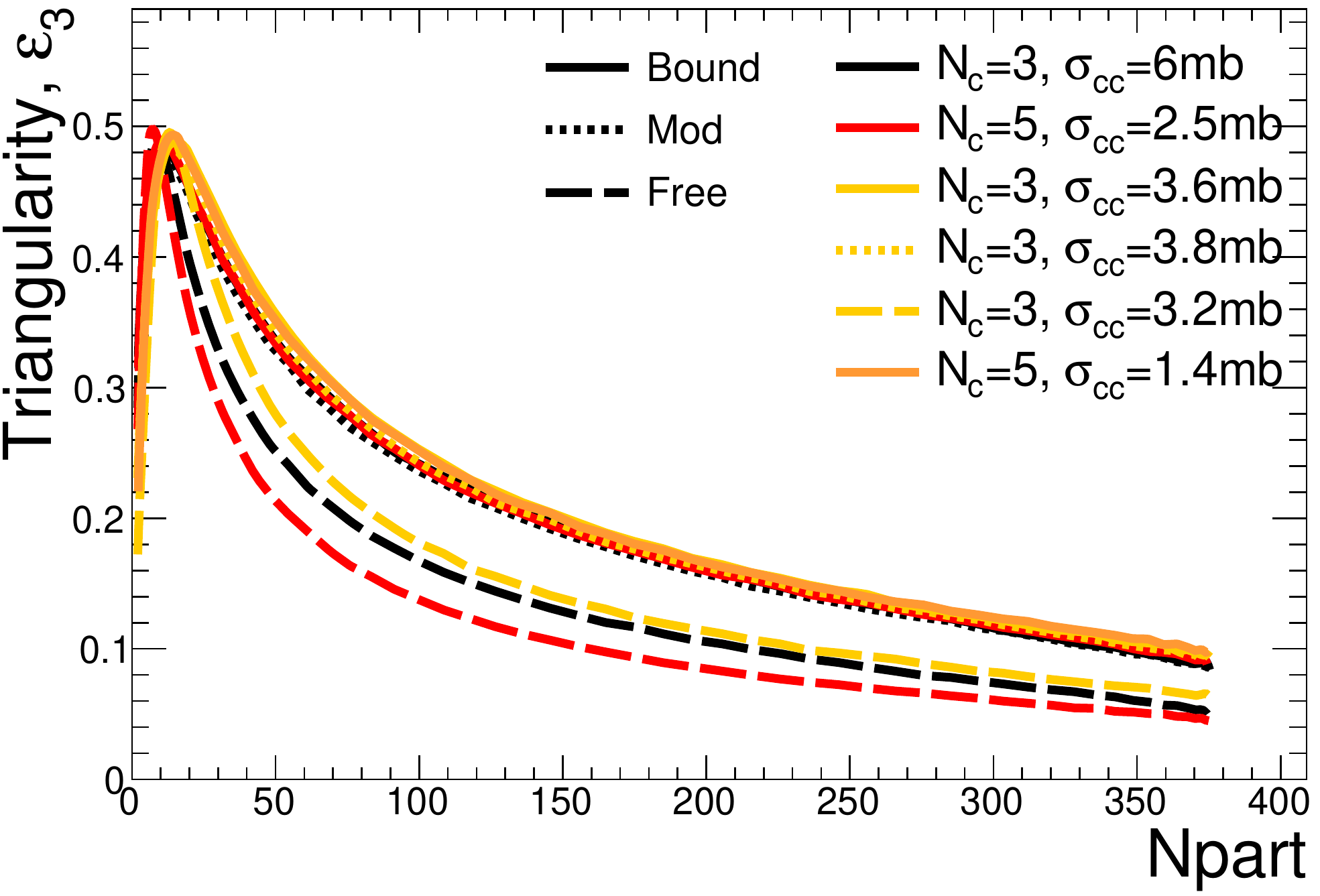}
   \includegraphics[width=0.235\textwidth]{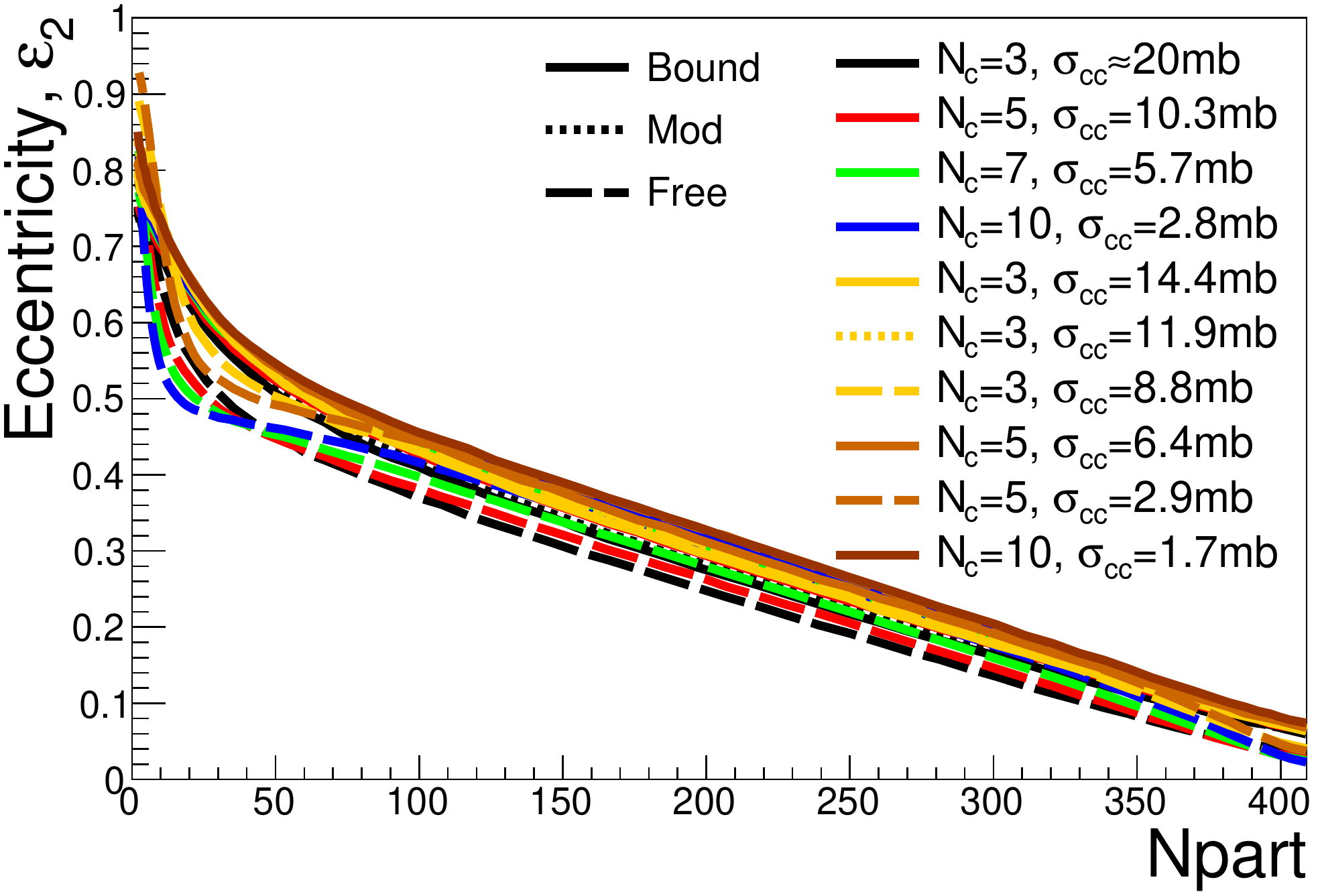}
   \includegraphics[width=0.235\textwidth]{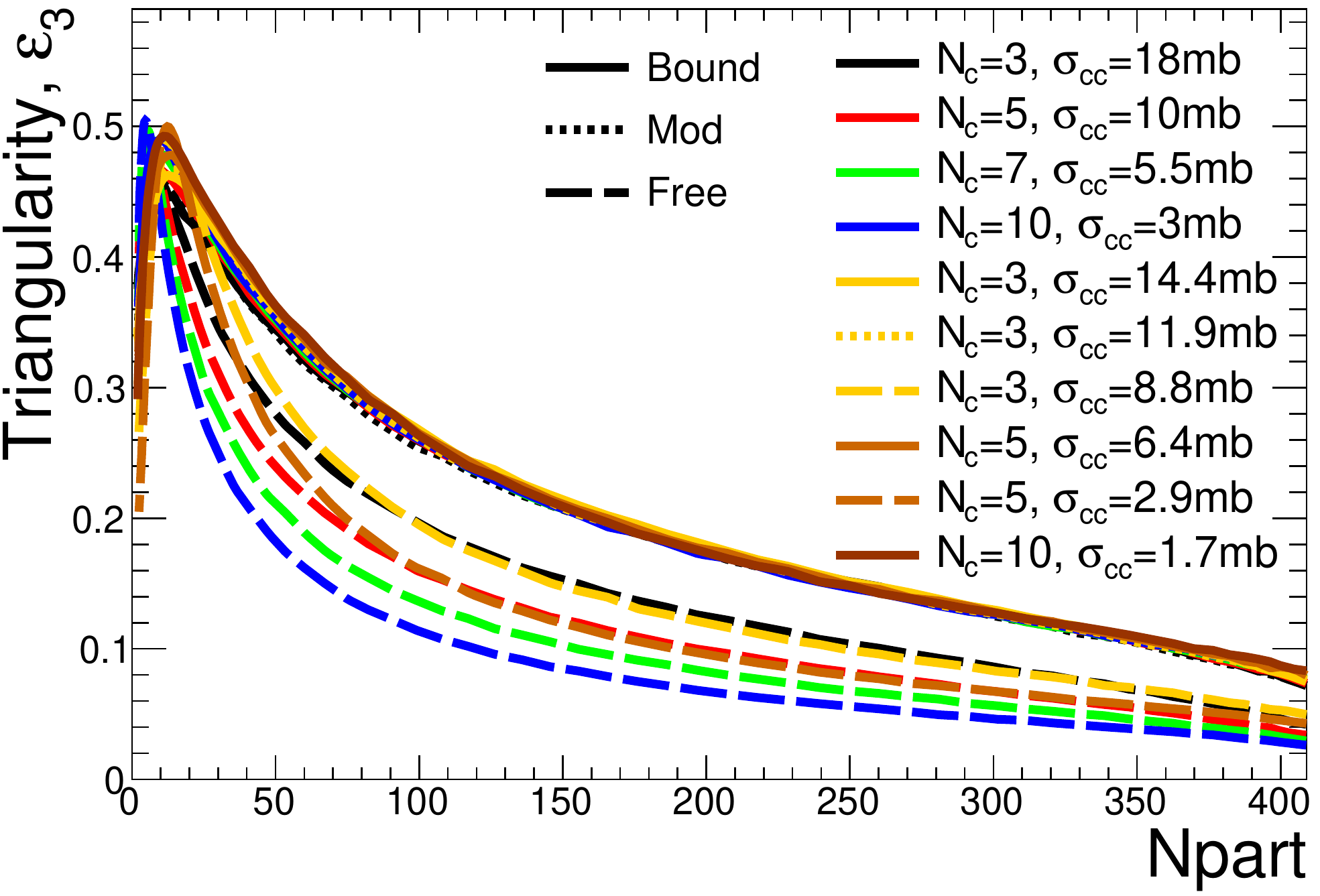}
   \caption{\label{fig:eccnuc}Eccentricity (left) and triangularity (right panels) for AuAu (top) and PbPb (bottom panels) collisions. The parameters for the calculations are summarized in \Tab{tab:pbpbvals} and \Tab{tab:auauvals}.}
\end{center}
\end{figure}

\Figure{fig:eccnuc} shows the eccentricity and triangularity versus $\Npart$ calculated for parameters given in \Tab{tab:pbpbvals} and \Tab{tab:auauvals}, which are quite similar to those calculated from participant nucleons, as also concluded in~\Ref{Miller:2003kd}.
The triangularity exhibits a stronger variation to changes of the calculation than the eccentricity, which is found to be quite insensitive to the actual values of the parameters.
As in the case of the nucleon participant calculation, $\varepsilon_{3}$ is only up to $10$--$20$\% larger than $\varepsilon_{2}$ in ultra-central collisions, which can not resolve the question why the measured $v_2\{2\}\approx v_3\{2\}$ in ultra-central collisions~\cite{Shen:2015qta,Shen:2015msa}.

\section{Summary}
\label{sec:sum}
Glauber models based on nucleon--nucleon interactions are commonly used to calculate properties of the initial state in high-energy nuclear collisions, and their dependence on impact parameter or number of participating nucleons.
Such calculations have be extended to the sub-nucleon level by taking into account three valence quarks per nucleon in the scattering process.
In particular, it has been shown that particle production at mid-rapidity in high-energy nucleus--nucleus collisions scales almost linearly with the number of quark participants.
In this paper, an extension to the Glauber model is presented, which accounts for an arbitrary number of effective sub-nucleon degrees of freedom, or partonic constituents, in the nucleons.
Properties of the initial state, such as the number of constituent participants and collisions, as well as eccentricity and triangularity, are calculated and systematically compared for different assumptions to distribute the sub-nuclear degrees of freedom and for various collision systems.
It is demonstrated that at high collision energy the number of produced particles scales with an average number of sub-nucleon degrees of freedom of between $3$ and $5$. 
As in the case of the nucleon participant calculation, $\varepsilon_{3}$ is only up to $10$--$20$\% larger than $\varepsilon_{2}$ in ultra-central collisions, which can not resolve the question why the measured $v_2\{2\}\approx v_3\{2\}$ in ultra-central collisions.
The code for the constituent Monte Carlo Glauber program is made publicly available.
The author welcomes comments on the code and suggestions on how to make it more useful to both experimentalists and theorists.

\section*{Acknowledgments}
I would like to thank J.Schukraft and S.Sorensen for interesting discussions. This work is supported in part by the U.S. Department of Energy, 
Office of Science, Office of Nuclear Physics, under contract number DE-AC02-05CH11231.

\newpage
\bibliography{biblio}{}

\begin{thebibliography}{49}
\expandafter\ifx\csname natexlab\endcsname\relax\def\natexlab#1{#1}\fi
\expandafter\ifx\csname bibnamefont\endcsname\relax
  \def\bibnamefont#1{#1}\fi
\expandafter\ifx\csname bibfnamefont\endcsname\relax
  \def\bibfnamefont#1{#1}\fi
\expandafter\ifx\csname citenamefont\endcsname\relax
  \def\citenamefont#1{#1}\fi
\expandafter\ifx\csname url\endcsname\relax
  \def\url#1{\texttt{#1}}\fi
\expandafter\ifx\csname urlprefix\endcsname\relax\def\urlprefix{URL }\fi
\providecommand{\bibinfo}[2]{#2}
\providecommand{\eprint}[2][]{\url{#2}}

\bibitem[{\citenamefont{Miller et~al.}(2007)\citenamefont{Miller, Reygers,
  Sanders, and Steinberg}}]{Miller:2007ri}
\bibinfo{author}{\bibfnamefont{M.~L.} \bibnamefont{Miller}},
  \bibinfo{author}{\bibfnamefont{K.}~\bibnamefont{Reygers}},
  \bibinfo{author}{\bibfnamefont{S.~J.} \bibnamefont{Sanders}},
  \bibnamefont{and}
  \bibinfo{author}{\bibfnamefont{P.}~\bibnamefont{Steinberg}},
  \bibinfo{journal}{Ann. Rev. Nucl. Part. Sci.} \textbf{\bibinfo{volume}{57}},
  \bibinfo{pages}{205} (\bibinfo{year}{2007}),
  \eprint{\arxiv{nucl-ex/0701025}}.

\bibitem[{\citenamefont{Bialas et~al.}(1976)\citenamefont{Bialas, Bleszynski,
  and Czyz}}]{Bialas:1976ed}
\bibinfo{author}{\bibfnamefont{A.}~\bibnamefont{Bialas}},
  \bibinfo{author}{\bibfnamefont{M.}~\bibnamefont{Bleszynski}},
  \bibnamefont{and} \bibinfo{author}{\bibfnamefont{W.}~\bibnamefont{Czyz}},
  \bibinfo{journal}{Nucl. Phys.} \textbf{\bibinfo{volume}{B111}},
  \bibinfo{pages}{461} (\bibinfo{year}{1976}).

\bibitem[{\citenamefont{Eskola et~al.}(1989)\citenamefont{Eskola, Kajantie, and
  Lindfors}}]{Eskola:1988yh}
\bibinfo{author}{\bibfnamefont{K.~J.} \bibnamefont{Eskola}},
  \bibinfo{author}{\bibfnamefont{K.}~\bibnamefont{Kajantie}}, \bibnamefont{and}
  \bibinfo{author}{\bibfnamefont{J.}~\bibnamefont{Lindfors}},
  \bibinfo{journal}{Nucl. Phys.} \textbf{\bibinfo{volume}{B323}},
  \bibinfo{pages}{37} (\bibinfo{year}{1989}).

\bibitem[{\citenamefont{Alver et~al.}(2008{\natexlab{a}})\citenamefont{Alver,
  Baker, Loizides, and Steinberg}}]{Alver:2008aq}
\bibinfo{author}{\bibfnamefont{B.}~\bibnamefont{Alver}},
  \bibinfo{author}{\bibfnamefont{M.}~\bibnamefont{Baker}},
  \bibinfo{author}{\bibfnamefont{C.}~\bibnamefont{Loizides}}, \bibnamefont{and}
  \bibinfo{author}{\bibfnamefont{P.}~\bibnamefont{Steinberg}}
  (\bibinfo{year}{2008}{\natexlab{a}}), \eprint{\arxiv{0805.4411}}.

\bibitem[{\citenamefont{Rybczynski et~al.}(2014)\citenamefont{Rybczynski,
  Stefanek, Broniowski, and Bozek}}]{Rybczynski:2013yba}
\bibinfo{author}{\bibfnamefont{M.}~\bibnamefont{Rybczynski}},
  \bibinfo{author}{\bibfnamefont{G.}~\bibnamefont{Stefanek}},
  \bibinfo{author}{\bibfnamefont{W.}~\bibnamefont{Broniowski}},
  \bibnamefont{and} \bibinfo{author}{\bibfnamefont{P.}~\bibnamefont{Bozek}},
  \bibinfo{journal}{Comput. Phys. Commun.} \textbf{\bibinfo{volume}{185}},
  \bibinfo{pages}{1759} (\bibinfo{year}{2014}), \eprint{\arxiv{1310.5475}}.

\bibitem[{\citenamefont{Eremin and Voloshin}(2003)}]{Eremin:2003qn}
\bibinfo{author}{\bibfnamefont{S.}~\bibnamefont{Eremin}} \bibnamefont{and}
  \bibinfo{author}{\bibfnamefont{S.}~\bibnamefont{Voloshin}},
  \bibinfo{journal}{Phys. Rev.} \textbf{\bibinfo{volume}{C67}},
  \bibinfo{pages}{064905} (\bibinfo{year}{2003}),
  \eprint{\arxiv{nucl-th/0302071}}.

\bibitem[{\citenamefont{Nouicer}(2007)}]{Nouicer:2006pr}
\bibinfo{author}{\bibfnamefont{R.}~\bibnamefont{Nouicer}},
  \bibinfo{journal}{Eur. Phys. J.} \textbf{\bibinfo{volume}{C49}},
  \bibinfo{pages}{281} (\bibinfo{year}{2007}),
  \eprint{\arxiv{nucl-th/0608038}}.

\bibitem[{\citenamefont{Adler et~al.}(2014)}]{Adler:2013aqf}
\bibinfo{author}{\bibfnamefont{S.~S.} \bibnamefont{Adler}} \bibnamefont{et~al.}
  (\bibinfo{collaboration}{PHENIX}), \bibinfo{journal}{Phys. Rev.}
  \textbf{\bibinfo{volume}{C89}}, \bibinfo{pages}{044905}
  (\bibinfo{year}{2014}), \eprint{\arxiv{1312.6676}}.

\bibitem[{\citenamefont{Adare et~al.}(2016)}]{Adare:2015bua}
\bibinfo{author}{\bibfnamefont{A.}~\bibnamefont{Adare}} \bibnamefont{et~al.}
  (\bibinfo{collaboration}{PHENIX}), \bibinfo{journal}{Phys. Rev.}
  \textbf{\bibinfo{volume}{C93}}, \bibinfo{pages}{024901}
  (\bibinfo{year}{2016}), \eprint{\arxiv{1509.06727}}.

\bibitem[{\citenamefont{Lacey et~al.}(2016)\citenamefont{Lacey, Liu, Magdy,
  Csanád, Schweid, Ajitanand, Alexander, and Pak}}]{Lacey:2016hqy}
\bibinfo{author}{\bibfnamefont{R.~A.} \bibnamefont{Lacey}},
  \bibinfo{author}{\bibfnamefont{P.}~\bibnamefont{Liu}},
  \bibinfo{author}{\bibfnamefont{N.}~\bibnamefont{Magdy}},
  \bibinfo{author}{\bibfnamefont{M.}~\bibnamefont{Csanád}},
  \bibinfo{author}{\bibfnamefont{B.}~\bibnamefont{Schweid}},
  \bibinfo{author}{\bibfnamefont{N.~N.} \bibnamefont{Ajitanand}},
  \bibinfo{author}{\bibfnamefont{J.}~\bibnamefont{Alexander}},
  \bibnamefont{and} \bibinfo{author}{\bibfnamefont{R.}~\bibnamefont{Pak}}
  (\bibinfo{year}{2016}), \eprint{\arxiv{1601.06001}}.

\bibitem[{\citenamefont{Zheng and Yin}(2016)}]{zheng}
\bibinfo{author}{\bibfnamefont{L.}~\bibnamefont{Zheng}} \bibnamefont{and}
  \bibinfo{author}{\bibfnamefont{Z.}~\bibnamefont{Yin}}, \bibinfo{journal}{Eur.
  Phys. J.} \textbf{\bibinfo{volume}{A52}}, \bibinfo{pages}{45}
  (\bibinfo{year}{2016}), \eprint{\arxiv{1603.02515}}.

\bibitem[{\citenamefont{Loizides}(2016)}]{Loizides:2016tew}
\bibinfo{author}{\bibfnamefont{C.}~\bibnamefont{Loizides}}
  (\bibinfo{year}{2016}), \eprint{\arxiv{1602.09138}}.

\bibitem[{\citenamefont{Bożek et~al.}(2016)\citenamefont{Bożek, Broniowski,
  and Rybczyński}}]{Bozek:2016kpf}
\bibinfo{author}{\bibfnamefont{P.}~\bibnamefont{Bożek}},
  \bibinfo{author}{\bibfnamefont{W.}~\bibnamefont{Broniowski}},
  \bibnamefont{and}
  \bibinfo{author}{\bibfnamefont{M.}~\bibnamefont{Rybczyński}},
  \bibinfo{journal}{Phys. Rev.} \textbf{\bibinfo{volume}{C94}},
  \bibinfo{pages}{014902} (\bibinfo{year}{2016}), \eprint{\arxiv{1604.07697}}.

\bibitem[{\citenamefont{Alver et~al.}(2008{\natexlab{b}})}]{Alver:2008zza}
\bibinfo{author}{\bibfnamefont{B.}~\bibnamefont{Alver}} \bibnamefont{et~al.},
  \bibinfo{journal}{Phys. Rev.} \textbf{\bibinfo{volume}{C77}},
  \bibinfo{pages}{014906} (\bibinfo{year}{2008}{\natexlab{b}}),
  \eprint{\arxiv{0711.3724}}.

\bibitem[{\citenamefont{De~Vries et~al.}(1987)\citenamefont{De~Vries, De~Jager,
  and De~Vries}}]{DeJager:1987qc}
\bibinfo{author}{\bibfnamefont{H.}~\bibnamefont{De~Vries}},
  \bibinfo{author}{\bibfnamefont{C.~W.} \bibnamefont{De~Jager}},
  \bibnamefont{and} \bibinfo{author}{\bibfnamefont{C.}~\bibnamefont{De~Vries}},
  \bibinfo{journal}{Atom. Data Nucl. Data Tabl.} \textbf{\bibinfo{volume}{36}},
  \bibinfo{pages}{495} (\bibinfo{year}{1987}).

\bibitem[{\citenamefont{Shou et~al.}(2015)\citenamefont{Shou, Ma, Sorensen,
  Tang, Videbæk, and Wang}}]{Shou:2014eya}
\bibinfo{author}{\bibfnamefont{Q.~Y.} \bibnamefont{Shou}},
  \bibinfo{author}{\bibfnamefont{Y.~G.} \bibnamefont{Ma}},
  \bibinfo{author}{\bibfnamefont{P.}~\bibnamefont{Sorensen}},
  \bibinfo{author}{\bibfnamefont{A.~H.} \bibnamefont{Tang}},
  \bibinfo{author}{\bibfnamefont{F.}~\bibnamefont{Videbæk}}, \bibnamefont{and}
  \bibinfo{author}{\bibfnamefont{H.}~\bibnamefont{Wang}},
  \bibinfo{journal}{Phys. Lett.} \textbf{\bibinfo{volume}{B749}},
  \bibinfo{pages}{215} (\bibinfo{year}{2015}), \eprint{\arxiv{1409.8375}}.

\bibitem[{\citenamefont{Loizides et~al.}(2015)\citenamefont{Loizides, Nagle,
  and Steinberg}}]{Loizides:2014vua}
\bibinfo{author}{\bibfnamefont{C.}~\bibnamefont{Loizides}},
  \bibinfo{author}{\bibfnamefont{J.}~\bibnamefont{Nagle}}, \bibnamefont{and}
  \bibinfo{author}{\bibfnamefont{P.}~\bibnamefont{Steinberg}},
  \bibinfo{journal}{SoftwareX} \textbf{\bibinfo{volume}{1-2}},
  \bibinfo{pages}{13} (\bibinfo{year}{2015}), \eprint{\arxiv{1408.2549}}.

\bibitem[{\citenamefont{Olive et~al.}(2014)}]{Agashe:2014kda}
\bibinfo{author}{\bibfnamefont{K.~A.} \bibnamefont{Olive}} \bibnamefont{et~al.}
  (\bibinfo{collaboration}{Particle Data Group}), \bibinfo{journal}{Chin.
  Phys.} \textbf{\bibinfo{volume}{C38}}, \bibinfo{pages}{090001}
  (\bibinfo{year}{2014}).

\bibitem[{\citenamefont{Aad et~al.}(2011)}]{Aad:2011eu}
\bibinfo{author}{\bibfnamefont{G.}~\bibnamefont{Aad}} \bibnamefont{et~al.}
  (\bibinfo{collaboration}{ATLAS}), \bibinfo{journal}{Nature Commun.}
  \textbf{\bibinfo{volume}{2}}, \bibinfo{pages}{463} (\bibinfo{year}{2011}),
  \eprint{\arxiv{1104.0326}}.

\bibitem[{\citenamefont{Antchev et~al.}(2011)}]{Antchev:2011vs}
\bibinfo{author}{\bibfnamefont{G.}~\bibnamefont{Antchev}} \bibnamefont{et~al.}
  (\bibinfo{collaboration}{TOTEM}), \bibinfo{journal}{Europhys. Lett.}
  \textbf{\bibinfo{volume}{96}}, \bibinfo{pages}{21002} (\bibinfo{year}{2011}),
  \eprint{\arxiv{1110.1395}}.

\bibitem[{\citenamefont{Chatrchyan et~al.}(2013)}]{Chatrchyan:2012nj}
\bibinfo{author}{\bibfnamefont{S.}~\bibnamefont{Chatrchyan}}
  \bibnamefont{et~al.} (\bibinfo{collaboration}{CMS}), \bibinfo{journal}{Phys.
  Lett.} \textbf{\bibinfo{volume}{B722}}, \bibinfo{pages}{5}
  (\bibinfo{year}{2013}), \eprint{\arxiv{1210.6718}}.

\bibitem[{\citenamefont{Cudell et~al.}(2002)\citenamefont{Cudell, Ezhela,
  Gauron, Kang, Kuyanov, Lugovsky, Martynov, Nicolescu, Razuvaev, and
  Tkachenko}}]{Cudell:2002xe}
\bibinfo{author}{\bibfnamefont{J.~R.} \bibnamefont{Cudell}},
  \bibinfo{author}{\bibfnamefont{V.~V.} \bibnamefont{Ezhela}},
  \bibinfo{author}{\bibfnamefont{P.}~\bibnamefont{Gauron}},
  \bibinfo{author}{\bibfnamefont{K.}~\bibnamefont{Kang}},
  \bibinfo{author}{\bibfnamefont{{\relax Yu}.~V.} \bibnamefont{Kuyanov}},
  \bibinfo{author}{\bibfnamefont{S.~B.} \bibnamefont{Lugovsky}},
  \bibinfo{author}{\bibfnamefont{E.}~\bibnamefont{Martynov}},
  \bibinfo{author}{\bibfnamefont{B.}~\bibnamefont{Nicolescu}},
  \bibinfo{author}{\bibfnamefont{E.~A.} \bibnamefont{Razuvaev}},
  \bibnamefont{and} \bibinfo{author}{\bibfnamefont{N.~P.}
  \bibnamefont{Tkachenko}} (\bibinfo{collaboration}{COMPETE}),
  \bibinfo{journal}{Phys. Rev. Lett.} \textbf{\bibinfo{volume}{89}},
  \bibinfo{pages}{201801} (\bibinfo{year}{2002}),
  \eprint{\arxiv{hep-ph/0206172}}.

\bibitem[{\citenamefont{ATLAS}(2015)}]{ATLAS-CONF-2015-038}
\bibinfo{author}{\bibnamefont{ATLAS}},
  \bibinfo{journal}{\href{https://cds.cern.ch/record/2045064}{ATLAS-CONF-2015-038}}
   (\bibinfo{year}{2015}).

\bibitem[{\citenamefont{CMS}(2015)}]{CMS:2016ael}
\bibinfo{author}{\bibnamefont{CMS}},
  \bibinfo{journal}{\href{https://cds.cern.ch/record/2145896?ln=en}{CMS-PAS-FSQ-15-005}}
   (\bibinfo{year}{2015}).

\bibitem[{\citenamefont{{D. d'Enterria and Klaus Reygers}}(2010)}]{dde}
\bibinfo{author}{\bibnamefont{{D. d'Enterria and Klaus Reygers}}},
  \bibinfo{journal}{\hrefurl{https://twiki.cern.ch/twiki/bin/view/Main/LHCGlauberBaseline}}
   (\bibinfo{year}{2010}).

\bibitem[{\citenamefont{Abelev et~al.}(2013)}]{Abelev:2013qoq}
\bibinfo{author}{\bibfnamefont{B.}~\bibnamefont{Abelev}} \bibnamefont{et~al.}
  (\bibinfo{collaboration}{ALICE}), \bibinfo{journal}{Phys. Rev.}
  \textbf{\bibinfo{volume}{C88}}, \bibinfo{pages}{044909}
  (\bibinfo{year}{2013}), \eprint{\arxiv{1301.4361}}.

\bibitem[{\citenamefont{Abelev et~al.}(2012)}]{ALICE:2012aa}
\bibinfo{author}{\bibfnamefont{B.}~\bibnamefont{Abelev}} \bibnamefont{et~al.}
  (\bibinfo{collaboration}{ALICE}), \bibinfo{journal}{Phys. Rev. Lett.}
  \textbf{\bibinfo{volume}{109}}, \bibinfo{pages}{252302}
  (\bibinfo{year}{2012}), \eprint{\arxiv{1203.2436}}.

\bibitem[{\citenamefont{Khachatryan et~al.}(2015)}]{Khachatryan:2015zaa}
\bibinfo{author}{\bibfnamefont{V.}~\bibnamefont{Khachatryan}}
  \bibnamefont{et~al.} (\bibinfo{collaboration}{CMS}) (\bibinfo{year}{2015}),
  \eprint{\arxiv{1509.03893}}.

\bibitem[{\citenamefont{Alver et~al.}(2011)}]{Alver:2010ck}
\bibinfo{author}{\bibfnamefont{B.}~\bibnamefont{Alver}} \bibnamefont{et~al.}
  (\bibinfo{collaboration}{PHOBOS}), \bibinfo{journal}{Phys. Rev.}
  \textbf{\bibinfo{volume}{C83}}, \bibinfo{pages}{024913}
  (\bibinfo{year}{2011}), \eprint{\arxiv{1011.1940}}.

\bibitem[{\citenamefont{Aggarwal et~al.}(2000)}]{Aggarwal:2000th}
\bibinfo{author}{\bibfnamefont{M.~M.} \bibnamefont{Aggarwal}}
  \bibnamefont{et~al.} (\bibinfo{collaboration}{WA98}), \bibinfo{journal}{Phys.
  Rev. Lett.} \textbf{\bibinfo{volume}{85}}, \bibinfo{pages}{3595}
  (\bibinfo{year}{2000}), \eprint{\arxiv{nucl-ex/0006008}}.

\bibitem[{\citenamefont{Adler et~al.}(2005{\natexlab{a}})}]{Adler:2005ig}
\bibinfo{author}{\bibfnamefont{S.~S.} \bibnamefont{Adler}} \bibnamefont{et~al.}
  (\bibinfo{collaboration}{PHENIX}), \bibinfo{journal}{Phys. Rev. Lett.}
  \textbf{\bibinfo{volume}{94}}, \bibinfo{pages}{232301}
  (\bibinfo{year}{2005}{\natexlab{a}}), \eprint{\arxiv{nucl-ex/0503003}}.

\bibitem[{\citenamefont{Chatrchyan et~al.}(2012)}]{Chatrchyan:2012vq}
\bibinfo{author}{\bibfnamefont{S.}~\bibnamefont{Chatrchyan}}
  \bibnamefont{et~al.} (\bibinfo{collaboration}{CMS}), \bibinfo{journal}{Phys.
  Lett.} \textbf{\bibinfo{volume}{B710}}, \bibinfo{pages}{256}
  (\bibinfo{year}{2012}), \eprint{\arxiv{1201.3093}}.

\bibitem[{\citenamefont{Alver et~al.}(2007)}]{Alver:2006wh}
\bibinfo{author}{\bibfnamefont{B.}~\bibnamefont{Alver}} \bibnamefont{et~al.}
  (\bibinfo{collaboration}{PHOBOS}), \bibinfo{journal}{Phys. Rev. Lett.}
  \textbf{\bibinfo{volume}{98}}, \bibinfo{pages}{242302}
  (\bibinfo{year}{2007}), \eprint{\arxiv{nucl-ex/0610037}}.

\bibitem[{\citenamefont{Alver and Roland}(2010)}]{Alver:2010gr}
\bibinfo{author}{\bibfnamefont{B.}~\bibnamefont{Alver}} \bibnamefont{and}
  \bibinfo{author}{\bibfnamefont{G.}~\bibnamefont{Roland}},
  \bibinfo{journal}{Phys. Rev.} \textbf{\bibinfo{volume}{C81}},
  \bibinfo{pages}{054905} (\bibinfo{year}{2010}), \bibinfo{note}{[Erratum:
  Phys. Rev.C82,039903(2010)]}, \eprint{\arxiv{1003.0194}}.

\bibitem[{\citenamefont{Teaney and Yan}(2011)}]{Teaney:2010vd}
\bibinfo{author}{\bibfnamefont{D.}~\bibnamefont{Teaney}} \bibnamefont{and}
  \bibinfo{author}{\bibfnamefont{L.}~\bibnamefont{Yan}},
  \bibinfo{journal}{Phys. Rev.} \textbf{\bibinfo{volume}{C83}},
  \bibinfo{pages}{064904} (\bibinfo{year}{2011}), \eprint{\arxiv{1010.1876}}.

\bibitem[{\citenamefont{d'Enterria et~al.}(2010)\citenamefont{d'Enterria,
  Eyyubova, Korotkikh, Lokhtin, Petrushanko, Sarycheva, and
  Snigirev}}]{d'Enterria:2010hd}
\bibinfo{author}{\bibfnamefont{D.}~\bibnamefont{d'Enterria}},
  \bibinfo{author}{\bibfnamefont{G.~K.} \bibnamefont{Eyyubova}},
  \bibinfo{author}{\bibfnamefont{V.~L.} \bibnamefont{Korotkikh}},
  \bibinfo{author}{\bibfnamefont{I.~P.} \bibnamefont{Lokhtin}},
  \bibinfo{author}{\bibfnamefont{S.~V.} \bibnamefont{Petrushanko}},
  \bibinfo{author}{\bibfnamefont{L.~I.} \bibnamefont{Sarycheva}},
  \bibnamefont{and} \bibinfo{author}{\bibfnamefont{A.~M.}
  \bibnamefont{Snigirev}}, \bibinfo{journal}{Eur. Phys. J.}
  \textbf{\bibinfo{volume}{C66}}, \bibinfo{pages}{173} (\bibinfo{year}{2010}),
  \eprint{\arxiv{0910.3029}}.

\bibitem[{\citenamefont{Hofstadter}(1956)}]{Hofstadter:1956qs}
\bibinfo{author}{\bibfnamefont{R.}~\bibnamefont{Hofstadter}},
  \bibinfo{journal}{Rev. Mod. Phys.} \textbf{\bibinfo{volume}{28}},
  \bibinfo{pages}{214} (\bibinfo{year}{1956}).

\bibitem[{\citenamefont{Mitchell et~al.}(2016)\citenamefont{Mitchell,
  Perepelitsa, Tannenbaum, and Stankus}}]{Mitchell:2016jio}
\bibinfo{author}{\bibfnamefont{J.~T.} \bibnamefont{Mitchell}},
  \bibinfo{author}{\bibfnamefont{D.~V.} \bibnamefont{Perepelitsa}},
  \bibinfo{author}{\bibfnamefont{M.~J.} \bibnamefont{Tannenbaum}},
  \bibnamefont{and} \bibinfo{author}{\bibfnamefont{P.~W.}
  \bibnamefont{Stankus}}, \bibinfo{journal}{Phys. Rev.}
  \textbf{\bibinfo{volume}{C93}}, \bibinfo{pages}{054910}
  (\bibinfo{year}{2016}), \eprint{\arxiv{1603.08836}}.

\bibitem[{\citenamefont{Aad et~al.}(2016)}]{Aad:2015gqa}
\bibinfo{author}{\bibfnamefont{G.}~\bibnamefont{Aad}} \bibnamefont{et~al.}
  (\bibinfo{collaboration}{ATLAS}), \bibinfo{journal}{Phys. Rev. Lett.}
  \textbf{\bibinfo{volume}{116}}, \bibinfo{pages}{172301}
  (\bibinfo{year}{2016}), \eprint{\arxiv{1509.04776}}.

\bibitem[{\citenamefont{Khachatryan et~al.}(2016)}]{Khachatryan:2016txc}
\bibinfo{author}{\bibfnamefont{V.}~\bibnamefont{Khachatryan}}
  \bibnamefont{et~al.} (\bibinfo{collaboration}{CMS}),
  \bibinfo{journal}{submitted to Physics Letters B}  (\bibinfo{year}{2016}),
  \eprint{\arxiv{1606.06198}}.

\bibitem[{\citenamefont{Corke and Sjostrand}(2011)}]{Corke:2011yy}
\bibinfo{author}{\bibfnamefont{R.}~\bibnamefont{Corke}} \bibnamefont{and}
  \bibinfo{author}{\bibfnamefont{T.}~\bibnamefont{Sjostrand}},
  \bibinfo{journal}{JHEP} \textbf{\bibinfo{volume}{05}}, \bibinfo{pages}{009}
  (\bibinfo{year}{2011}), \eprint{1101.5953}.

\bibitem[{\citenamefont{Adler et~al.}(2005{\natexlab{b}})}]{Adler:2004zn}
\bibinfo{author}{\bibfnamefont{S.~S.} \bibnamefont{Adler}} \bibnamefont{et~al.}
  (\bibinfo{collaboration}{PHENIX}), \bibinfo{journal}{Phys. Rev.}
  \textbf{\bibinfo{volume}{C71}}, \bibinfo{pages}{034908}
  (\bibinfo{year}{2005}{\natexlab{b}}), \bibinfo{note}{[Erratum: Phys.
  Rev.C71,049901(2005)]}, \eprint{\arxiv{nucl-ex/0409015}}.

\bibitem[{\citenamefont{Adam et~al.}(2015)}]{Adam:2015ptt}
\bibinfo{author}{\bibfnamefont{J.}~\bibnamefont{Adam}} \bibnamefont{et~al.}
  (\bibinfo{collaboration}{ALICE}) (\bibinfo{year}{2015}),
  \eprint{\arxiv{1512.06104}}.

\bibitem[{\citenamefont{Aamodt et~al.}(2011)}]{Aamodt:2010cz}
\bibinfo{author}{\bibfnamefont{K.}~\bibnamefont{Aamodt}} \bibnamefont{et~al.}
  (\bibinfo{collaboration}{ALICE}), \bibinfo{journal}{Phys. Rev. Lett.}
  \textbf{\bibinfo{volume}{106}}, \bibinfo{pages}{032301}
  (\bibinfo{year}{2011}), \eprint{\arxiv{1012.1657}}.

\bibitem[{\citenamefont{Miller and Snellings}(2003)}]{Miller:2003kd}
\bibinfo{author}{\bibfnamefont{M.}~\bibnamefont{Miller}} \bibnamefont{and}
  \bibinfo{author}{\bibfnamefont{R.}~\bibnamefont{Snellings}}
  (\bibinfo{year}{2003}), \eprint{\arxiv{nucl-ex/0312008}}.

\bibitem[{\citenamefont{Shen et~al.}(2015)\citenamefont{Shen, Qiu, and
  Heinz}}]{Shen:2015qta}
\bibinfo{author}{\bibfnamefont{C.}~\bibnamefont{Shen}},
  \bibinfo{author}{\bibfnamefont{Z.}~\bibnamefont{Qiu}}, \bibnamefont{and}
  \bibinfo{author}{\bibfnamefont{U.}~\bibnamefont{Heinz}},
  \bibinfo{journal}{Phys. Rev.} \textbf{\bibinfo{volume}{C92}},
  \bibinfo{pages}{014901} (\bibinfo{year}{2015}), \eprint{1502.04636}.

\bibitem[{\citenamefont{Shen and Heinz}(2015)}]{Shen:2015msa}
\bibinfo{author}{\bibfnamefont{C.}~\bibnamefont{Shen}} \bibnamefont{and}
  \bibinfo{author}{\bibfnamefont{U.}~\bibnamefont{Heinz}},
  \bibinfo{journal}{\arxiv{1507.01558}}  (\bibinfo{year}{2015}).

\bibitem[{\citenamefont{Brun and Rademakers}(1997)}]{Brun:1997pa}
\bibinfo{author}{\bibfnamefont{R.}~\bibnamefont{Brun}} \bibnamefont{and}
  \bibinfo{author}{\bibfnamefont{F.}~\bibnamefont{Rademakers}},
  \bibinfo{journal}{Nucl. Instrum. Meth.} \textbf{\bibinfo{volume}{A389}},
  \bibinfo{pages}{81} (\bibinfo{year}{1997}).

\bibitem[{\citenamefont{Bjorken}(1983)}]{Bjorken:1982qr}
\bibinfo{author}{\bibfnamefont{J.~D.} \bibnamefont{Bjorken}},
  \bibinfo{journal}{Phys. Rev.} \textbf{\bibinfo{volume}{D27}},
  \bibinfo{pages}{140} (\bibinfo{year}{1983}).

\end{thebibliography}
\appendix
\section{Program code}
\label{sec:code}
The program code, called ``runCGM.C'', for the generalized constituent Monte Carlo Glauber can be found at
\hrefurl{http://tglaubermc.hepforge.org/svn/branches/tools/runCGM.C}.
It requires ``runglauber\_v2.3.C'' from the most recent TGlauberMC version (v2.3)~\cite{Loizides:2014vua}, which can be downloaded from from HepForge (\hrefurl{http://www.hepforge.org/downloads/tglaubermc}), and ROOT~\cite{Brun:1997pa} (see \hrefurl{http://root.cern.ch} for installation files and documentation.).
To compile the code, execute at the ROOT prompt:
\begin{verbatim}
 .L runglauber_2.3.C+
 .L runCGM.C+
\end{verbatim}
\ifcode
The function ``runCGM'' can be run with the following arguments:
\begin{verbatim}
Int_t n           = number of events
const char *sysA  = system A
const char *sysB  = system B
Double_t signn    = NN cross section (mb)
Double_t mind     = min. dist. betw. nucleons
Int_t nc          = number of constituents / dof
Double_t sigcc    = constituent cross section (mb)
Int_t type,       -> how to distribute dof:
                      =0 no recentering
                      =5 modfied (PHENIX)
                      =8 free no recentering 
const char *fname = output filename 
Double_t bmin     = min. imp. parameter
Double_t bmax     = max. imp. parameter
\end{verbatim}
The output ROOT ``ntuple'' contains the following list of per-event variables:
\begin{verbatim}
Npart  = number of nucleon participants
Ncoll  = number of nucleon collisions
B      = impact parameter
Ncpart = number of constituent participants
Nccoll = number of constituent collisions
Ap     = area def. by participant (co-)variances
Ac     = area def. by constituent (co-)variances
EccXP  = eccX nucleon participants (X=1-5)
EccXC  = eccX constituent participants (X=1-5)
\end{verbatim}
\else
See the source code for options on how to run the program and the description of the output ROOT ``ntuple''.
\fi
All distributions discussed in \Sec{sec:results} have been obtained from the output of ``runGCM''.
\ifcomment
\fi

\begin{figure}[th!]
\begin{center}
   \includegraphics[width=0.44\textwidth]{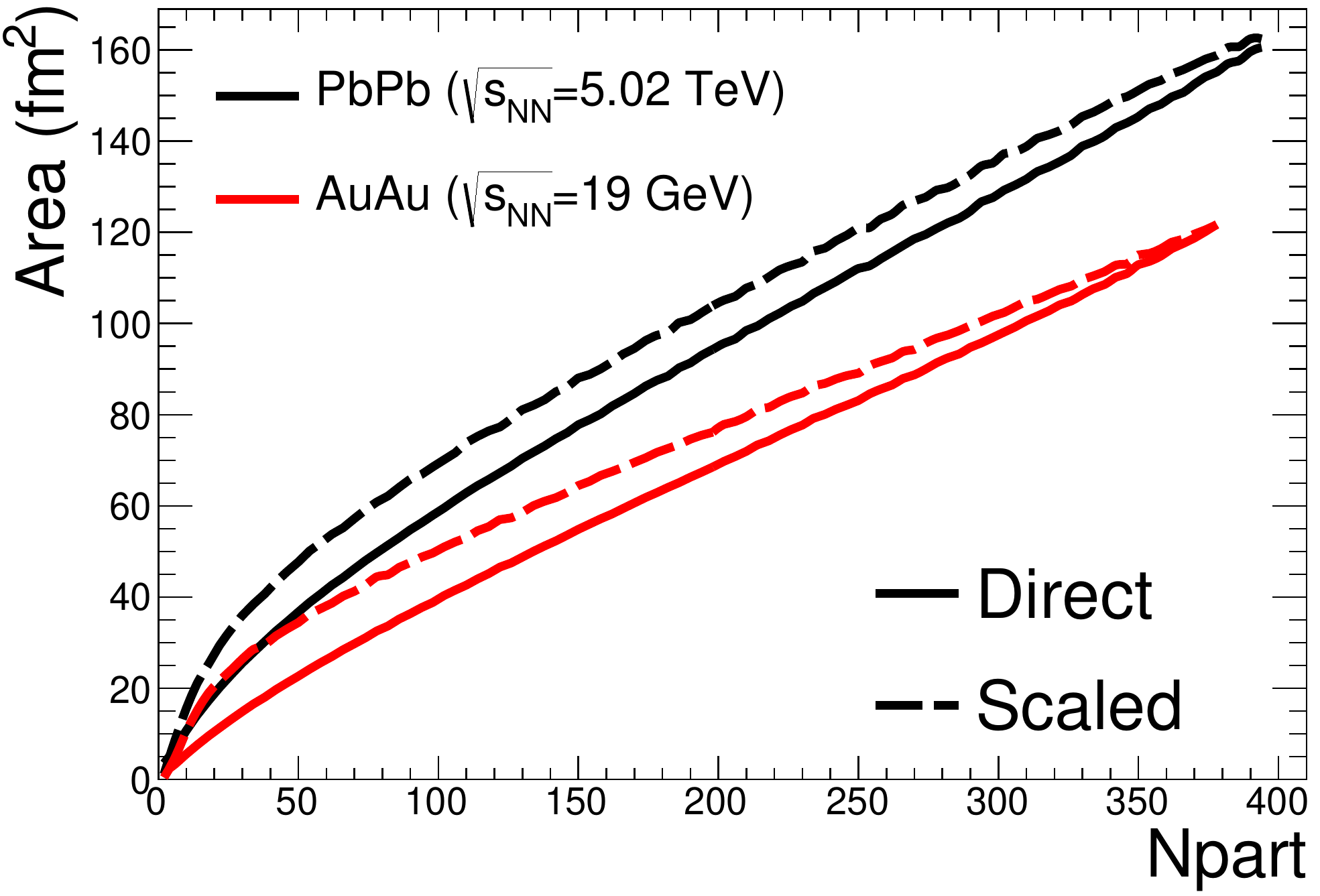}
   \caption{\label{fig:areadirect}Area calculated directly by counting of the overlap area and scaled using the participant widths for AuAu collisions at $\snn=19$ GeV and PbPb collisions at $\snn=5.02$ TeV.}
\end{center}
\end{figure}
\begin{figure}[th!]
\begin{center}
   \includegraphics[width=0.44\textwidth]{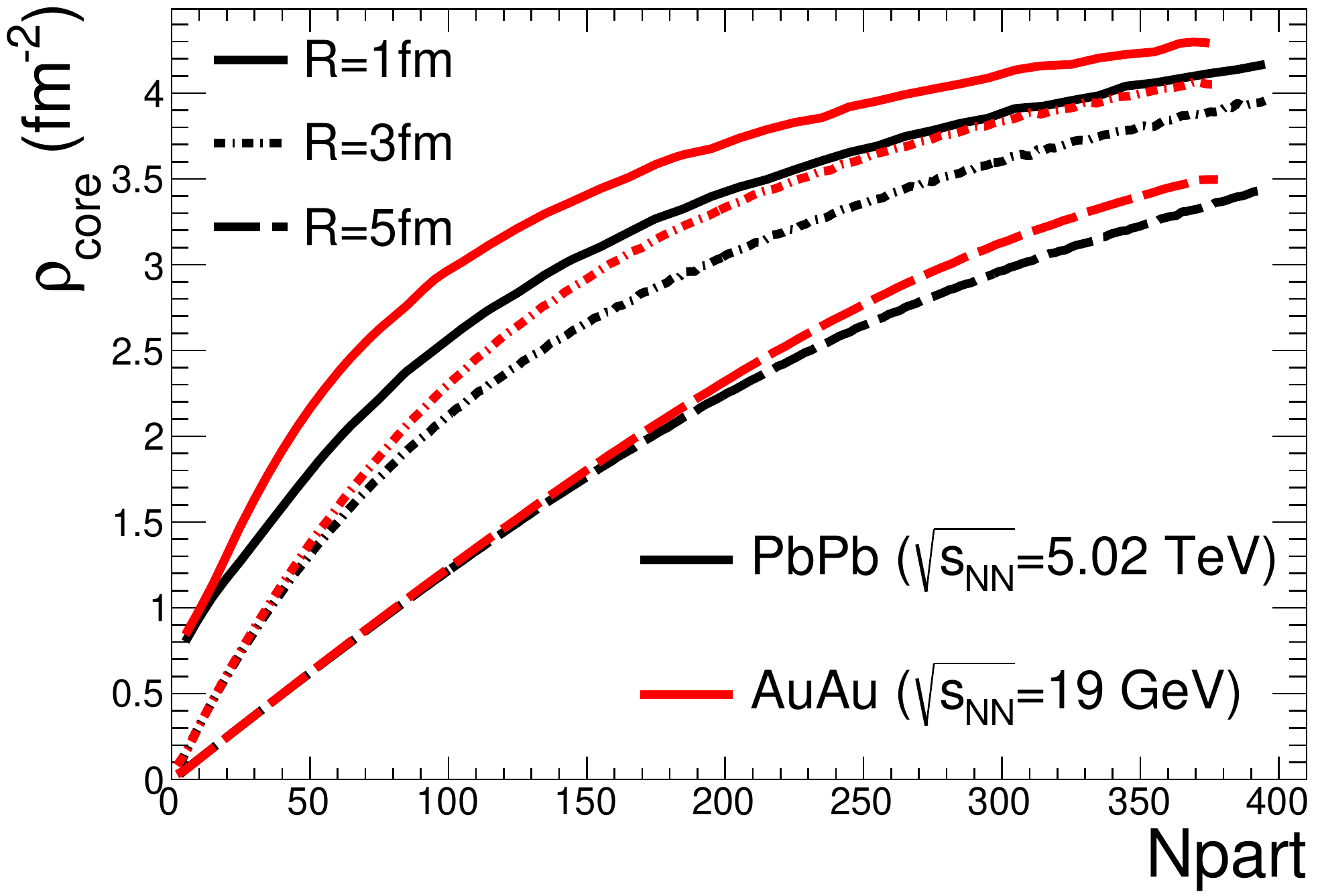}
   \caption{\label{fig:partdens}Participant transverse area density in an area given by radius $R=1$, $3$ and $5$~fm for AuAu collisions at $\snn=19$ GeV and PbPb collisions at $\snn=5.02$ TeV.}
\end{center}
\end{figure}

\section{Area calculation}
\label{sec:area}
As briefly mentioned in \Sec{sec:calculation}, the overlap area of two colliding nuclei is usually taken to be proportional to $S=\sqrt{\sigma^2_{x}\sigma^2_{y}-\sigma^2_{xy}}$, given by the (co-)variances of the participant distributions in the transverse plane~\cite{Alver:2008zza}.
However, using the participant distributions does not provide a direct measure of the area, and in particular misses also the absolute normalization.
Instead, one can event-by-event compute the overlap area directly using a fine-grained grid.
\Figure{fig:areadirect} compares the two approaches for AuAu collisions at $\snn=19$ GeV and PbPb collisions at $\snn=5.02$ TeV, where the results using the participant widths were rescaled by $A_0/S_0$ where $S_0$ and the absolute area $A_0$ were obtained at $b=0$~fm.
The values are $S_0=9.8$ and $8.7$ with RMS of $0.4$, and $A_0=165.8$ and $120.1$ with RMS of $5.2$ and $3.9$ for PbPb and AuAu, respectively (all units in fm$^2$).   
The code can be found at \hrefurl{http://tglaubermc.hepforge.org/svn/branches/tools/runArea.C}.

Alternatively, instead of directly using the area when estimating the energy density via the Bjorken estimate~\cite{Bjorken:1982qr}, one can use the participant transverse area density, $\rho_{\rm core}$, which can be obtained by counting the number of participants within a core area of given radius $R$.
\Figure{fig:partdens} shows $\rho_{\rm core}$ for various choices of $R$ in AuAu collisions at $\snn=19$ GeV and PbPb collisions at $\snn=5.02$ TeV.
The code can be found at \hrefurl{http://tglaubermc.hepforge.org/svn/branches/tools/runCore.C}

\ifextra
\section{Extra}
\begin{figure}[t]
\begin{center}
   \includegraphics[width=0.235\textwidth]{cnccollpbpb}\hspace{0.125cm}
   \includegraphics[width=0.235\textwidth]{cncpartpbpb}
    \includegraphics[width=0.49\textwidth]{cncpbynpartpbpb}
   \caption{\label{fig:e1}}
\end{center}
\end{figure}

\begin{figure}[t]
\begin{center}
   \includegraphics[width=0.235\textwidth]{cnccollauau}\hspace{0.125cm}
   \includegraphics[width=0.235\textwidth]{cncpartauau}
    \includegraphics[width=0.49\textwidth]{cncbynpartauau}
   \caption{\label{fig:e2}}
\end{center}
\end{figure}

\begin{figure}[ht!]
\begin{center}
   \includegraphics[width=0.45\textwidth]{cbarea}
   \caption{\label{fig:e3}Area calculated directly from the MC and via S for AuAu collisions at $\snn=19$ GeV and PbPb collisions at $\snn=5.02$ TeV.}
\end{center}
\end{figure}
\fi
\end{document}